\documentclass[format=acmsmall, review=false, screen=true]{acmart}
\usepackage{booktabs} %
\usepackage{threeparttable}
\usepackage{subfigure}
\usepackage[linesnumbered, ruled]{algorithm2e}

\SetAlFnt{\small}
\SetAlCapFnt{\small}
\SetAlCapNameFnt{\small}
\SetAlCapHSkip{0pt}
\IncMargin{-\parindent}

\acmJournal{TORS}
\usepackage{amsmath,amsfonts}
\usepackage{algorithmic}
\usepackage{graphicx}
\usepackage{xcolor}
\usepackage{hyperref}
\usepackage{setspace}
\usepackage{booktabs} 
\usepackage{subfigure}
\usepackage{amsmath}
\usepackage{color}
\usepackage{amsmath}
\usepackage{mathrsfs}
\usepackage[normalem]{ulem}
\usepackage{enumitem}
\usepackage{multirow}
\usepackage{ulem}
\usepackage[nomargin,inline,marginclue,draft]{fixme}
\usepackage{balance}
\usepackage{verbatim}
\usepackage{diagbox}
\usepackage{changepage}
\usepackage{xcolor}
\usepackage{bm}
\usepackage{multicol}

\acmJournal{TOIS}

\newcommand{\ie}{\emph{i.e., }}
\newcommand{\eg}{\emph{e.g., }}
\newcommand{\etal}{\emph{et al. }}

\newcommand{\wrt}{\emph{w.r.t. }}
\newcommand{\cf}{\emph{cf. }}
\newcommand{\aka}{\emph{a.k.a. }}
\usepackage[justification=centering]{caption}
\usepackage{caption}
\definecolor{myred}{rgb}{1, 0, 0}
\definecolor{myblue}{rgb}{0, 0, 1}
\definecolor{myblack}{rgb}{1, 1, 1}

 \newcommand{\work}[1]{}

\newcommand{\para}[1]{{\vspace{4pt} \bf \noindent #1 \hspace{0pt}}}

\newcommand{\revise}[1]{\textcolor{black}{#1}}

\newlength\savedwidth

\newcommand{\delinarxiv}[1]{#1}

\begin{document}

\title{Causal Inference in Recommender Systems:\\ A Survey and Future Directions}

\author{Chen Gao}
\affiliation{%
		\institution{Beijing National Research Center for Information Science and Technology, Tsinghua University}
		\country{China}}
\author{Yu Zheng}
\affiliation{%
		\institution{Department of Electronic Engineering, Tsinghua University}
		\country{China}}
\author{Wenjie Wang}
\affiliation{%
		\institution{School of Computing, National University of Singapore}
		\country{Singapore}}
	\email{chgao96@gmail.com, y-zheng19@mails.tsinghua.edu.cn, wenjiewang96@gmail.com, fulifeng93@gmail.com, xiangnanhe@gmail.com, liyong07@tsinghua.edu.cn}
\author{Fuli Feng}
\affiliation{%
		\institution{School of Information Science and Technology, University of Science and Technology of China}
		\country{China}}		
\author{Xiangnan He}
\affiliation{%
		\institution{School of Information Science and Technology, University of Science and Technology of China}
		\country{China}}
\author{Yong Li}
\affiliation{%
		\institution{Department of Electronic Engineering, Tsinghua University}
		\country{China}}
						
	\renewcommand{\shortauthors}{Gao~\textit{et al.}}

\begin{abstract}
Recommender systems have become crucial in information filtering nowadays. Existing recommender systems extract user preferences based on the correlation in data, such as behavioral correlation in collaborative filtering, feature-feature, or feature-behavior correlation in click-through rate prediction. However, unfortunately, the real world is driven by \textit{causality}, not just correlation, and correlation does not imply causation. For instance, recommender systems might recommend a battery charger to a user after buying a phone, where the latter can serve as the cause of the former; such a causal relation cannot be reversed. Recently, to address this, researchers in recommender systems have begun utilizing causal inference to extract causality, thereby enhancing the recommender system. In this survey, we offer a comprehensive review of the literature on causal inference-based recommendation. Initially, we introduce the fundamental concepts of both recommender system and causal inference as the foundation for subsequent content. We then highlight the typical issues faced by non-causality recommender system. Following that, we thoroughly review the existing work on causal inference-based recommender systems, based on a taxonomy of three-aspect challenges that causal inference can address. Finally, we discuss the open problems in this critical research area and suggest important potential future works.

\end{abstract}


\maketitle

\section{Introduction}\label{sec:introduction}

In the era of information overload, recommender systems (RecSys) have emerged as the fundamental service for facilitating users' information access.
From the early shallow models~\cite{koren2009matrix,rendle2009bpr} to recent advances of deep learning-based ones~\cite{he2017neural,covington2016deep} and the most recent graph neural network-based models~\cite{ying2018graph,he2020lightgcn}, the techniques and models of recommender systems are developing rapidly.
In general, recommender systems aim to learn user preferences by fitting historical behaviors, along with collected user profiles, item attributes, or other contextual information. Here, the interaction is mainly induced by the previous recommender system and is largely affected by the recommendation policy.
Then, recommender systems filter from the item-candidate pools and select items that match users' personalized preferences and demands. Once deployed, the system collects new interactions to update the model, where the whole framework thus forms a feedback loop.

Generally, recommender systems can be divided into two categories: collaborative filtering (CF) and content-based recommendation (\textit{a.k.a.}, click-through rate (CTR) prediction, shortened as CTR prediction).
Collaborative filtering focuses on users' historical behaviors, such as clicking, purchasing, etc.
The basic assumption of collaborative filtering is that users with similar historical behaviors tend to have similar future behaviors.
For example, the most representative matrix factorization model (MF) uses vectors to represent users and items, and then it uses the inner product to calculate the relevance scores between users and items.
To improve the model capacity, recent work~\cite{he2017neural,covington2016deep} takes advantage of deep neural networks for matching users with items, such as neural collaborative filtering~\cite{he2017neural}, which leverages multi-layer perceptrons to replace the inner product in the MF model.
Furthermore, a broad view of collaborative filtering models the relevance with consideration of additional information, such as
the timestamp of each behavior in sequential recommendation~\cite{chen2022intent,zhang2022dynamic}, user social network in social recommendation~\cite{fan2019graph,wu2019neural}, and multi-type behaviors in multi-behavior recommendation~\cite{gao2019neural,xia2021graph}, etc.
CTR prediction focuses on leveraging the rich attributes and features of users, items, or context to enhance recommendation. The mainstream CTR prediction task aims to learn high-order features with the proper feature-interaction module, such as the linear inner product in Factorization Machine (FM), multi-layer perceptrons in DeepFM~\cite{guo2017deepfm}, attention networks in AFM~\cite{xiao2017atteational}, stacked self-attention layers in AutoInt~\cite{song2019autoint}, etc.

The basis of today's recommender systems is to model the \textit{correlation}, such as behavioral correlation in collaborative filtering, feature-feature, or feature-behavior correlation in click-through rate prediction.
However, the real world is driven by \textit{causality} rather than correlation, while correlation does not imply causation.
Two kinds of causality widely exist in recommender systems, user-aspect, and interaction-aspect.
The user-aspect causality refers to the users' decision process being driven by causality.
For example, a user may buy a battery charger after buying a phone, in which the latter can serve as the cause of the former, and such a causal relation cannot be reversed.
The interaction-aspect causality refers to that the recommendation strategy largely affects users' interactions with the system.
For example, the unobserved user-item interaction does not mean that the user does not like the item, which may only be caused by non-exposure.

Formally speaking, causality can be defined as \textit{cause} and \textit{effect} in which the cause is partly responsible for the effect~\cite{yao2020survey}. Causal inference is defined as the process of determining and further leveraging the causal relation based on experimental data or observational data~\cite{yao2020survey}. 
Two popular and widely-used causal-inference frameworks are the potential outcome framework (Rubin Causal Model)~\cite{rubin1974estimating}, and the structural causal model (SCM)~\cite{pearl1995causal,pearl2009causality}.
Rubin's Framework aims to calculate the effect of certain treatments.
The structural causal model establishes a causal graph and corresponding structural equations, comprising a set of variables and structural equations that depict the causal relationships between these variables.

Since following a correlation-driven paradigm, existing recommender systems still suffer from critical bottlenecks. Specifically, three main challenges limit the effectiveness of the current paradigm, for which causal inference can serve as a promising solution, as follows.
\begin{itemize}[leftmargin=*]
    \item \textbf{The issues of data bias}. The collected data, such as the most important user-item interaction data, is observational (not experimental), resulting in biases including conformity bias, popularity bias, etc.~\cite{lin2021mitigating} 
    As for the non-causality recommender systems, not only the desired user preferences but also the data bias are learned by the model, which leads to inferior recommendation performance.
    \item \textbf{The issues of data missing or even data noise.} 
    The collected data in recommender systems is limited by the collection procedure, which makes there is missing or noisy data. For example, despite the large-scale item pool, the users only interact with a tiny fraction of items, which means plenty of unobserved user-item feedback cannot be collected. Moreover, sometimes the observed implicit feedback is even noisy, not reflecting the actual satisfaction of users, such as those click behaviors that end with negative reviews on E-Commerce websites or some behaviors by mistake.
    \item \textbf{The beyond-accuracy objectives are hard to achieve.}
    Besides accuracy, recommender systems should also consider other objectives, such as fairness, explainability, transparency, etc. Improving these beyond-accuracy objectives may hurt the recommendation accuracy, resulting in a dilemma. For example, a model that considers the \textit{multiple driven causes under user behavior}, based on assigning each cause with disentangled and interpretable embedding, can well provide both accurate and explainable recommendation. Another important objective is diversity but a high-diversity item recommendation list may not be able to well fit user interest. Here causal inference can help capture \textit{why users consume specific category} of items, achieving both high  accuracy and diversity. 
    
\end{itemize}

\begin{figure}[t]
\includegraphics[width=0.8\linewidth]{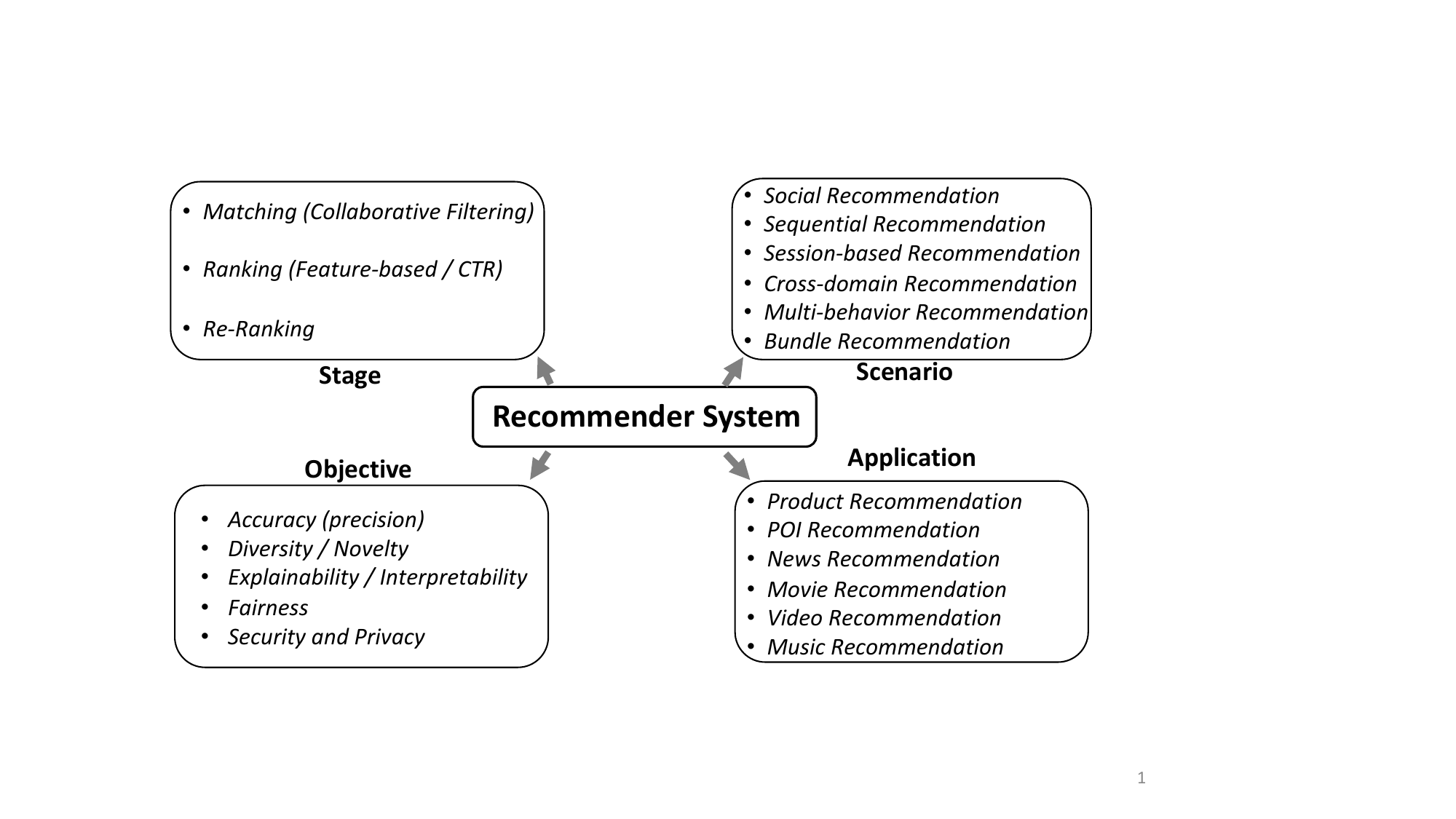} 
\caption{A simple comparison among three kinds of data issues: data bias, data missing, and data noise, taking collaborative filtering as an example. In general, data bias refers to the biased data collection (\textit{e.g.}, in conformity bias, user behavior does not full reflect preferences as it may be due to conformity); data missing refers to the unobserved preferences (labeled with question marks); data noise refers to incorrect data (marked with red color). As a simple illustration, this figure does not cover other recommendation tasks.} \label{fig:data-issues}
\end{figure}

Recent research on recommender systems tackles these challenges with carefully-designed causality-driven methods. Over the last two years, there has been a surge of relevant papers, and there is a very high probability that causal inference will become predominant in the field of recommender systems. In this survey paper, we systematically review these pioneering research efforts, especially focusing on how they address the critical shortcomings with causal inference.

First, recommendation methods incorporating causality can construct a causal graph. Within this framework, bias is typically viewed as a confounder, which can then be addressed using causal-inference techniques. Second, regarding the issue of missing data, causality-enhanced models can assist in constructing a counterfactual world. Thus, the missing data can be inferred through counterfactual reasoning. Third, causal inference naturally facilitates the development of interpretable and controllable models. As a result, the explainability of both the model itself and the recommendation outcomes can be enhanced. Moreover, other objectives, such as diversity and fairness, can also be realized since the model becomes more controllable. Specifically, the current works of causal inference in recommendation can be categorized as follows.
\begin{itemize}[leftmargin=*]
\item \textbf{Data debiasing with causal inference.} For issues like popularity bias or exposure bias, the bias (arising from popularity-aware or exposure strategy-aware data collection) can often be seen as a form of confounder. Some existing work addresses this through backdoor adjustment. Conformity bias, on the other hand, can be conceptualized as a collider effect. 
\item \textbf{Data augmentation and data denoising with causal inference.} The dual challenge of data missing encompasses both limited user-data collection and the recommendation model's causal effect on the system. The extreme form of the first challenge can even lead to data noise. 
For the first challenge, counterfactual reasoning can be employed to generate the uncollected data as augmentation, thus addressing the data-missing problem.
For the latter, causal models like IPW can be utilized to estimate the causal impact of recommendation models.
\item \textbf{Achieving explainability, diversity, and fairness via interpretable and controllable recommendation models using causal inference.} Models crafted in alignment with the causal graph are intrinsically controllable. Some notable techniques in this regard encompass causal discovery and disentangled representations. Leveraging the interpretable model, high diversity can be realized by manipulating the model to sidestep the tradeoff, and fair recommendations can be secured by steering the model to ensure fairness across specific user demographics.
\end{itemize}

It is worth mentioning that although there are surveys on either recommender systems~\cite{zhang2019deep,guo2020survey2,wu2022survey} or causal inference~\cite{guo2020survey,yao2021survey,moraffah2020causal,moraffah2020causal}, there is no existing survey fully discussing this new and important area of causality-driven recommender systems. \delinarxiv{Note that there is a very short paper (8 page)~\cite{wu2022opportunity} trying to survey existing work of causal-inspired recommendation methods, but it only discusses a few of representative papers due to its page limit.}
These surveys on recommender systems mainly introduce and discuss the basic concepts and various advances of recommender systems, with only a few discussions on causality-based recommendation.
On the other hand, surveys of causal inference primarily introduce and discuss the basic concepts and fundamental methods of causal inference, lacking sufficient discussions on applications.

	There is a survey~\cite{chen2023bias} about bias and debias in recommender system and we would discuss its relations with our survey as follows.  First, the survey~\cite{chen2023bias} concentrates on the bias issue in recommendations and describes how various works address these issues. Among these, causal inference-based methods represent just one segment, with numerous other methods available for tackling bias. Similarly, our survey underscores that while using causal inference to address data bias is a significant component, it is merely a portion of our broader theme: causal inference for recommender systems. Hence, even though some overlap exists between the two surveys, it is small due to the distinct focal topics. Second, when considering the shared part, the two surveys adopt different manners to discuss existing works. Our survey places greater emphasis on the causal inference technique itself, its ties to conventional causal inference methods, and its relevance to other challenges, such as data missing and data noise. In contrast, the bias survey~\cite{chen2023bias} delves deeper into the intricacies of biases (types, origins, etc.) and elaborates on how causal inference-based methods differentiate themselves from other kinds of methods.

We summarize the contribution of this survey as follows.
\begin{itemize}[leftmargin=*]
	\item To the best of our knowledge, we take the pioneering step to give a systematic survey of this new yet promising area. We categorize the existing work by answering the fundamental question of \textit{why the causal inference is needed and how causal inference enhances recommendation}. 
	\item  We first provide the necessary knowledge of recommender systems and causal inference. Subsequently, we introduce and explain the existing work of causal inference for recommendation, from the early attempts to the recently-published papers until 2023.
	\item We discuss important yet unresolved problems in this  research area and propose promising directions, which we believe will be the mainstream research direction of the next few years.
\end{itemize}

\section{Background}\label{sec::background}

\begin{figure*}[t]
	\centering
	\includegraphics[width=0.6\linewidth]{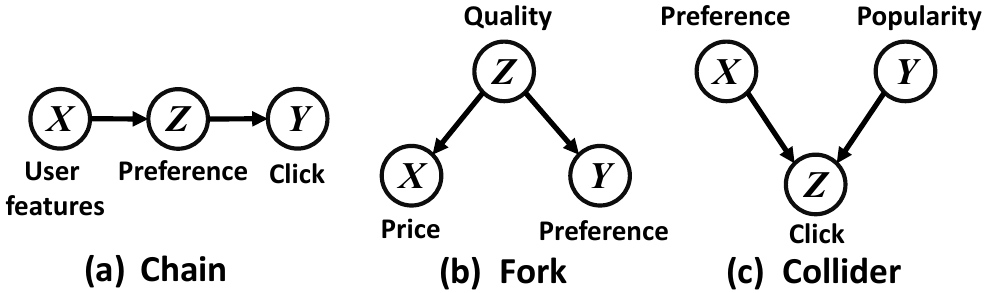} 
	\caption{Illustration of three typical DAGs.} \label{fig:DAGs}
\end{figure*}

As a survey of the interdisciplinary area of causal inference and recommender systems, we first introduce the background knowledge and fundamental concepts of these two topics.

\subsection{Causal Inference}
\begin{figure*}[t]
	\centering
	\includegraphics[width=0.8\linewidth]{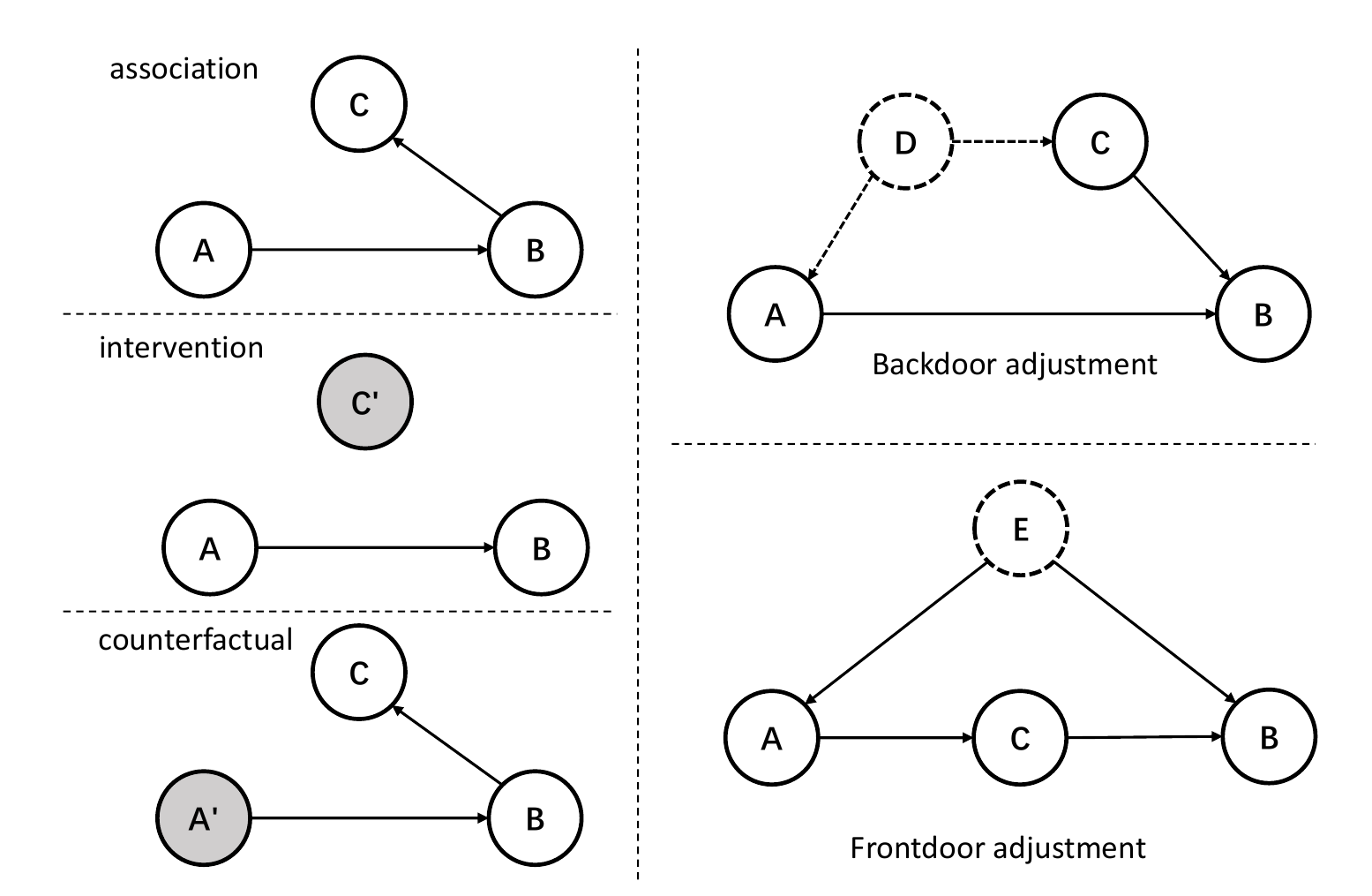} 
	\caption{Important concepts of causal inference.} \label{fig:causal-concepts}
\end{figure*}

We introduce the fundamental concepts of causal inference to facilitate the readers' understanding. This involves two representative causal frameworks: SCMs (Structural Causal Models) proposed by Pearl \textit{et al.}~\cite{pearl2009causality} and the potential outcome framework developed by Rubin \textit{et al.}~\cite{rubin1974estimating}. 
Considering the topic of this survey, we will elaborate on the core concepts using examples from recommender systems for clearer understanding. The basic concepts are shown in Fig. ~\ref{fig:DAGs} and  Fig.~\ref{fig:causal-concepts}, which we will explain in detail in the following sections.

\subsubsection{Structural Causal Models}
Generally, SCMs abstract the causal relationships between variables into causal graphs, build structural functions and then conduct causal inference to estimate the effects of interactions or counterfactuals~\cite{pearl2009causality}. 

\textbf{Causal Models.} 
Causal models involve two essential concepts: causal graphs and structural functions. 
Specifically, a causal graph describes the causal relationships via a Directed Acyclic Graph (DAG), in which the nodes denote variables and the edges indicate causal relationships. 
According to a causal graph, structural functions are used to model the relationships. For each variable, one structural function calculates its value based on its parent nodes.

\textbf{Three Typical DAGs.} 
As shown in Fig.~\ref{fig:DAGs}, there are three classic structures in causal graphs: \textit{chain}, \textit{fork}, and \textit{collider}, for each of which we give an example of recommender systems.
In the chain structure, $X$ affects $Y$ via the mediator $Z$. 
For example, in Fig~\ref{fig:DAGs} (a), the user features affect the user preferences, and the user preferences affect the users' click behavior.
Besides, in the fork structure, $Z$ is a confounder, affecting both $X$ and $Y$. 
For example, as shown in In Fig.~\ref{fig:DAGs}(b), an item's quality can affect both its price and users' preferences towards it. In such a fork structure, $Z$ is defined as \textit{confounder variable}.
Roughly ignoring confounder $Z$ leads to \textbf{spurious} correlation between $X$ and $Y$.
That is, products with higher prices may have larger sales on an e-commerce platform, which does not mean users prefer to spend much money.
In Fig.~\ref{fig:DAGs}(c), differently, $Z$ represents a collider, which is affected by $X$ and $Z$.
For example, the users' click behavior is affected by user preference and item popularity.
Conditioning on $Y$ will lead to \textbf{correct} correlation between $X$ and $Z$. 
That is, users' behaviors on two items with the same popularity level are only affected by their preferences.

\textbf{Intervention.}
Given the causal graph, a basic concept of intervention can be formally defined.
Specifically, the intervention on a variable $X$ is formulated with $do$-calculus, $do(X=x)$~\cite{pearl2009causality}, which blocks the effect of $X$'s parents and set the value of $X$ as $x$. 
For example, $do(X=x)$ in Fig.~\ref{fig:DAGs}(b) will rule out the path $Z\rightarrow X$ and force $X$ to be $x$~\cite{Pearl2018the}. That is, in our above-mentioned example, we set the item prices to a specific value.

\textbf{Counterfactual.} Another important concept is the counterfactual, which contrasts with the factual. It is used to address scenarios where the treatment variable's value settings do not occur in the real world.   In other words, counterfactual inference estimates what the outcome would have been if the treatment variable had taken on a different value compared to its observed value in the real world~\cite{pearl2009causality}. 
For example, a bankrupted seller might wonder about potential sales in a  counterfactual world where he/she had purchased advertisement services, setting the treatment variable  $T_{\textbf{if\_ads}}=1$.

\subsubsection{Potential Outcome Framework}
The potential outcome framework~\cite{rubin1974estimating} is another widely-used causal inference framework besides the structural causal model~\cite{pearl2009causality}.
It estimates the causal effect of a treatment variable on an outcome variable without the need for a causal graph.

\textbf{Potential Outcome~\cite{rubin1974estimating}.}
Given the treatment variable $T$ and the outcome variable $Y$, the potential outcome $Y_{t}^{i}$ denotes the value of $Y$ under the treatment $T=t$ for individual $i$. 
In the factual world, we can only observe the potential outcome of $Y$ under one treatment for each individual. 

\textbf{Treatment Effect~\cite{rubin1974estimating}.} 
Given binary treatments $T=0$ or $1$, the \textbf{Individual Treatment Effect} (ITE) for an individual $i$ is defined as $Y_{1}^{i} - Y_{0}^{i}$. 
However, ITE is impossible to calculate since we can only observe one potential outcome. Hence, ITE is extended to \textbf{Average Treatment Effect} (ATE) over a population.
For a population $i=\{1,2, ..., N\}$, ATE is calculated by $\mathbb{E}_i\left[Y_{1}^{i} - Y_{0}^{i}\right] = \frac{1}{N}\sum_{i=1}^{N}\left(Y_{1}^{i} - Y_{0}^{i}\right)$.

\noindent \textbf{Discussions about these two frameworks.}
We briefly summarize the similarities and differences between the two frameworks. As stated by Pearl~\cite{pearl2009causal}, the two frameworks are logically equivalent. The theorem and assumptions in one framework can be equivalently translated into the language of the other framework. However, the key difference is that the potential outcome framework neither considers the causal graph to describe causal relationships nor conducts reasoning over the graph to estimate causal effects.

\subsubsection{Causal Effect Estimation and Causal Discovery}

For estimating the causal effect, one golden rule is to conduct randomized experiments.
Since individuals are divided into the treatment group and the control group randomly, there are no unobserved confounders.
Under randomized experiments, some favorable properties of causal inference are guaranteed, such as covariate balance and exchangeability. 
Meanwhile, the causal effect can be estimated directly by comparing the two groups.
For example, online A/B testing can be regarded as a kind of randomized experiment that divides users randomly into several groups and can obtain trustworthy evaluation results of recommendation performance.

However, randomized experiments can be expensive and sometimes impossible to conduct.
For example, in recommender systems, experiments generating randomized recommendations can detrimentally affect user experiences and the platform's profitability. 
Therefore, estimating the causal effect solely from observational data becomes critical. 
In general, a causal estimand is first transformed into a statistical estimand with a causal model like SCM.
Then the statistical estimand is estimated with observed data.
In other words, with the defined causal model, we can discern causal effects and non-causal effects, such as confounding associations between treatment and outcome.
 Subsequently, the causal effect is extrapolated by estimation using observed data in alignment with the identified causal mechanisms.

One classical method is \textbf{backdoor adjustment}~\cite{pearl2009causality}.
We say a set of variables $W$ satisfies \textit{backdoor criterion}
if $W$ contain no descendant of T and $W$ can block backdoor paths (which has arrow into $T$ rather than from $T$) between $T$ and $Y$.
The causal effect of $T$ on $Y$ then can be obtained with backdoor adjustment as follows,
\begin{equation}
    P(y \mid \mathrm{do}(t)) = \sum_{w}{P(y \mid t,w)P(w)},
\end{equation}
where $w \in W$ and the total causal effect is the weighted sum of the conditioned causal effect.

The above backdoor adjustment can address observed confounders, but not unobserved confounders, where \textbf{frontdoor adjustment}~\cite{pearl2009causality} comes to help.
We say a set of variables $M$ satisfies \textit{frontdoor criterion} if all the causal paths from treatment variable $T$ to the outcome variable $Y$ are through $M$, and there is no unblocked backdoor path from $T$ to $M$, as well as $M$ to $Y$ when conditioned on $T$.
The causal effect of $T$ on $Y$ then can be obtained with frontdoor adjustment as follows,
\begin{equation}
    P(y \mid \mathrm{do}(t)) = \sum_{m}{P(m \mid t)\sum_{t'}{P(y \mid m, t')P(t')}},
\end{equation}
where possible unobserved confounders are addressed.

With the sufficient adjustment set of variables $W$ in the high dimension, it is difficult to directly estimate the causal effect as the positivity property is hard to satisfy.
Instead of modeling the whole set $W$, we can turn to the propensity score as follow, 
\begin{equation}
    e(W) = P(T=1 \mid W)
\end{equation}
which indicates the probability of receiving the treatment given $W$.
Then the causal effect can be estimated by inverse propensity weighting (\textbf{IPW})~\cite{hirano2003efficient} on the treatment and control group as follows,
\begin{equation}
    \hat{\tau} = \frac{1}{n_1}\sum_{i: t_i=1}{\frac{y_i}{e(w_i)}} - \frac{1}{n_2}\sum_{j: t_j=0}{\frac{y_j}{1-e(w_j)}}.
\end{equation}

All of the above causal effect estimations assume that we already have a causal graph. 
However, in the real world, we  often lack prior knowledge about the causal relationships in collected data.
This limitation gives rise to the problem of causal discovery, where the objective is to construct a causal graph from the existing data of a set of variables. 
Traditional approaches identify causal relations through conditional independence tests, bolstered by additional assumptions such as faithfulness~\cite{spirtes2000causation}. 
Score-based algorithms~\cite{heckerman1995learning,schwarz1978estimating} have been also proposed to relax the strict assumptions for causal discovery.
These methods utilize a score function to measure the quality of the discovered causal graph in comparison with observed data.
Recently, various machine learning approaches have been developed to discover causal relations from large-scale data.
For example, Zhu \textit{et al.}~\cite{zhu2019causal} utilize reinforcement learning method to find an optimal DAG with respect to a scoring function and penalties on acyclicity. There is a survey~\cite{guo2020survey} fully discusses different methods of causal discovery.

To summarize it, we have introduced the fundamental knowledge of causal inference, including two basic frameworks and two important research topics, causal effect estimation and causal discovery.

\subsection{Recommender System}

\subsubsection{Overview}
As an approach to information filtering, the recommender system has been widely deployed on various platforms in recent decades, such as TikTok, YouTube, Twitter, etc. 
In general, the modeling of user preferences based on historical interactions is the key point for the recommendation algorithm, and users' future interactions are further predicted. In this way, the necessary data input of a recommendation task includes the records of user-item interactions, and the output is a model that can generate the interaction likelihood of a given user-item pair. This procedure can be formulated as,
\begin{equation}
\begin{aligned}
    &\textbf{Input}: \mathbf{Y} \in \mathbb{R}^{\vert \mathcal{U}\vert \times \vert \mathcal{I} \vert}, \\
    &\textbf{Output}: f(\cdot, \cdot), (u, i) \xrightarrow{f} \mathbb{R},
\end{aligned}
\end{equation}
where $\mathcal{U}$ and $\mathcal{I}$ denotes the user set and item set, respectively. $y_{u', i'} = 1$ if user $u'\in \mathcal{U}$ has interacted with item $i'\in \mathcal{I}$; if not, $y_{u', i'} = 0$; here the function $f(\cdot, \cdot)$ denotes the recommendation model. 
Furthermore, with different input data, there are two primary families of models in recommendation, \textit{i.e.}, collaborative filtering (CF) and click-through rate (CTR) prediction.
Despite the vanilla CF which only considers user-item interaction data, some recommendation tasks enhance the behavioural data with auxiliary data, such as social network in social recommendation~\cite{wu2019diffnet, fan2019graphrec_social}, behavioral sequences in sequential recommendation~\cite{chang2021sequential, zhu2021graph}, multiple-type behaviors in multi-behavior recommendation~\cite{jin2020multi, zhang2020multiplex}, multi-domain user behaviors in cross-domain recommendation~\cite{hu2018conet, gao2019cross}, etc. 
For CTR prediction problem, user and item features such as user profiles~(occupation, age) and item attributes~(category, brand) are also considered as input.
The mainstream works of CTR prediction focus on extracting high-order cross-features with attention-based neural network~\cite{guo2017deepfm}, attention-based neural network~\cite{guo2017deepfm}, self-attentive layers~\cite{song2019autoint}, etc. 

\subsubsection{Recommendation Model Design}
Here we present two folds of design of recommendation models, collaborative filtering and click-through rate prediction.

\para{Collaborative Filtering.} Following the development process, existing CF models can be categorized into three types, including matrix factorization~(MF)-based, neural network~(NN)-based, and graph neural network~(GNN)-based. 
The standard way of modeling is to represent users and items with latent vectors, \textit{i.e.}, embeddings. 
With user embedding matrix $\mathbf{P}\in \mathbb{R}^{d \times \vert \mathcal{U}\vert}$ and item embedding matrix $\mathbf{Q}\in \mathbb{R}^{d \times \vert \mathcal{I}\vert}$, in which $d$ denotes embedding dimension, the interaction likelihood of $(u, i)$ will be the similarity of corresponding embeddings $\mathbf{p}_u$ and $\mathbf{q}_i$. 
\begin{itemize}
    \item \textbf{MF}~\cite{koren2009matrix}. The similarity function is the inner product as follows,
    \begin{equation}
        s(u, i) = \mathbf{p}_u^\top \mathbf{q}_i.
    \end{equation}
    
    \item \textbf{NCF}~\cite{he2017neural}. In order to incorporate the capability of modeling non-linearity, NCF generalized the similarity function and introduced the multi-layer perceptron~(MLP) as follows,
    \begin{equation}
    \begin{aligned}
        &s(u, i) = \mathbf{h}^\top\left(\mathbf{p}_u^G \odot \mathbf{q}_i^G\right) + \phi \left([\mathbf{p}_u^M, \mathbf{q}_i^M]\right), \\
        &\mathbf{p}_u = [\mathbf{p}_u^G, \mathbf{p}_u^M], \mathbf{q}_i = [\mathbf{q}_i^G, \mathbf{q}_i^M],
    \end{aligned}
    \end{equation}
    where $\mathbf{p}_u^G, \mathbf{p}_u^M$~($\mathbf{q}_i^G, \mathbf{q}_i^M$) denotes the user~(item) embedding for MF and MLP parts respectively, $[\cdot, \cdot]$ indicates the concatenation operation,  $\odot$ indicates the Hadamard product, $\mathbf{h}$ is the weight vector, and $\phi(\cdot)$ denotes MLP.
    
    \item \textbf{NGCF}~\cite{wang2019ngcf}. This GNN-based recommendation model conducts multiple layers of message passing on the user-item bipartite graph. Formally, the similarity is calculated as follows,
    \begin{equation}
    \begin{aligned}
        &\mathbf{p}_u^l = \textbf{Agg}\left(\mathbf{q}_i^{l-1} \vert i\in \mathcal{N}_u\right), \mathbf{q}_i^l = \textbf{Agg}\left(\mathbf{p}_u^{l-1} \vert u\in \mathcal{N}_i\right), \\
        &s(u, i) =\phi ( \left([\mathbf{p}_u^0, \cdots, \mathbf{p}_u^L]\right)^\top [\mathbf{q}_i^0, \cdots, \mathbf{q}_i^L]),
    \end{aligned}
    \end{equation}
    where $\mathbf{p}_u^0 = \mathbf{p}_u, \mathbf{q}_i^0 = \mathbf{q}_i$, and $l$ indicates the propagation layer, $\mathcal{N}_u$ refers to the set of interacted items of user $u$, and $\mathcal{N}_i$ indicates the set of those users who have interacted with item $i$. Here $\textbf{Agg}(\cdot)$ is the aggregation function for collecting neighborhood information. In this way, high-order user-item connectivity is injected into the similarity measurement between nodes.
\end{itemize}

\para{Click-Through Rate Prediction.} As introduced above, the unified procedure of CTR prediction is extracting high-order features. The input features are denoted as follows,
\begin{equation}
    \mathbf{x}_{u, i} = [\mathbf{x}_{u, i}^1, \cdots, \mathbf{x}_{u, i}^M],
\end{equation}
where $M$ denotes the number of feature fields. Furthermore, the raw features will be transformed into embeddings as follows,
\begin{equation}
    \mathbf{v}_{u, i}^k = \mathbf{V}^k\mathbf{x}_{u, i}^k, k = 1, \cdots, M,
\end{equation}
where $\mathbf{V}^k\in \mathbb{R}^{d^k\times \vert \mathcal{F}^k\vert}$ is the feature embedding matrix, $\mathcal{F}^k$ is the set of optional features, $d^k$ is the dimension of embeddings, and $k$ denotes the order of feature field. In general, there are two fields of users' and items' identity, supposed to be the first two ones, then $\mathbf{V}^1 = \mathbf{P}$ and $\mathbf{V}^2 = \mathbf{Q}$. In terms of the mapping function, it can be represented as follows,
\begin{equation}
    s(u, i) = g\left([\mathbf{v}_{u, i}^1, \cdots, \mathbf{v}_{u, i}^M]\right).
\end{equation}
The design of $g(\cdot)$ will introduce a module of feature interaction learning,  via the inner product in FM~\cite{rendle2010factorization}, multi-layer perceptions in DeepFM~\cite{guo2017deepfm}, stacked self-attention layers in AutoInt~\cite{song2019autoint}, etc.

\subsubsection{Objective Function}
The primary objective functions for optimization utilized in recommendation models are in two categories, \textit{i.e.}, point-wise and pair-wise. Specifically, the point-wise objective function focuses on the prediction of a user-item interaction of which the widely-used Logloss function is as follows,
\begin{equation}
    \mathcal{L} = -\frac{1}{\vert \mathcal{O}\vert}\sum_{(u, i)\in \mathcal{O}}y_{u, i}\log(\hat{y}_{u, i})+(1-y_{u, i})\log(1-\hat{y}_{u, i}),
\end{equation}
where $\hat{y}_{u, i} = s(u, i)$ and $\mathcal{O}$ is the training set. 

In terms of the pair-wise objective function, it encourages a larger disparity between positive~($y_{u, i}=1$) and negative~($y_{u, j}=0$) samples, and the widely-used BPR loss function~\cite{rendle2009bpr} is as follows,
\begin{equation}
    \mathcal{L} = -\frac{1}{\vert \mathcal{Q}\vert}\sum_{(u, i, j)\in \mathcal{Q}, \hat{y}_{u, i} > \hat{y}_{u, j}}\log\sigma(\hat{y}_{u, i} - \hat{y}_{u, j}),
\end{equation}
where $\sigma(\cdot)$ denotes the sigmoid function, and $\mathcal{Q}$ denotes the training set.
\section{Why Causal Inference is Needed for Recommender Systems}\label{sec::motivation}

\begin{figure}[t]
\centering
\includegraphics[width=0.8\linewidth]{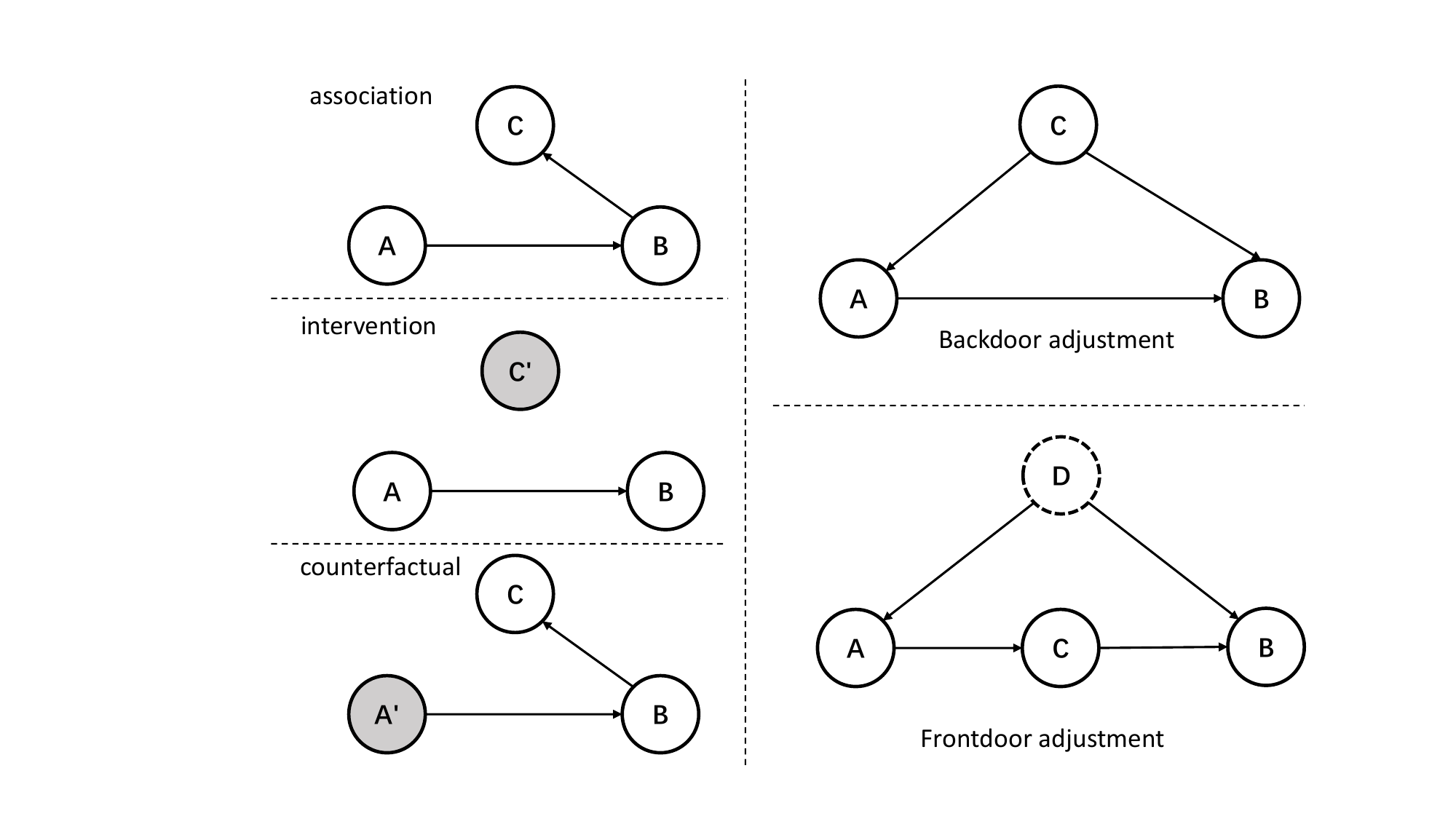} 
\caption{Illustration of three typical issues of non-causality recommendation models and how causal inference addresses them.} \label{fig:three-issues}
\end{figure}

In this section, we will discuss the essentiality and benefits of introducing causal inference into recommender systems from three aspects, illustrated in Fig.~\ref{fig:three-issues}.

\subsection{The Issues of Data Bias in Recommender Systems} 

\subsubsection{Data bias in recommender systems}

\textit{Data bias} refers to the uneven distribution of recommendation data that does not faithfully reflect user preference. Generally, there are two main types of data bias in recommendation over interactions and attributes.

\noindent\textbf{Bias over interactions.} Historical user-item interactions collected from previous recommendation strategies are typically treated as labels for recommender model training. 
Sometimes, historical interactions follow a highly skewed distribution over items (\aka long-tail distribution), resulting in models over-recommend popular items, \ie \textbf{popularity bias}~\cite{wei2021model,DBLP:conf/sigir/ZhangF0WSL021}. 
Furthermore, the historical interactions of a user also exhibit uneven distributions over item categories. Consequently, recommender models will blindly assign high scores to items from the frequent category, ignoring the user preference over the remaining categories~\cite{DBLP:conf/kdd/WangF0WC21}. 
Worse still, such biases will be amplified in the feedback loop, leading to notorious issues like unfairness and the filter bubble.
\textbf{Conformity bias} refers to the fact that users' behaviors are determined by not only user preferences but also conformity, making the collected data biased. It is a common issue in social-aware information systems, such as the user-post interaction behavior on Facebook\footnote{https://facebook.com}.
\textbf{Exposure bias} is another widely-concerned bias, which refers to that the exposure algorithms will highly influence the data collection of user feedback.

\noindent\textbf{Bias over attributes.} Item attributes that can directly result in interactions, especially clicks, can also mislead the estimation of user preference. Training over historical interactions will inevitably push the model to highlight such attributes, leading to shortcuts. 
Taking video recommendation as an example, videos with attractive titles or cover images are more likely to be clicked, while the user may not like the content~\cite{DBLP:conf/sigir/WangF0ZC21}.
Undeniably, the shortcuts of such item attributes will lead to recommendations failing to satisfy user preference. Worse still, they also make the recommender system vulnerable to relevant attacks, \eg the item producer intentionally leverages such features.

\subsubsection{The necessity of causal inference for data debiasing}
Causal theory enables us to identify the root cause of data bias by scrutinizing the generation procedure of recommendation data and mitigating the impact of bias through causal recommendation modeling.

\noindent\textbf{Causal view of data bias.} The main source of bias effect in recommendation is the backdoor path (Fig.~\ref{fig:DAGs}(b)), where a confounder ($Z$) simultaneously affects the inputs ($X$) and interactions ($Y$). Due to the existence of the backdoor path, directly estimating the correlation between $X$ and $Y$ will suffer from spurious correlations, leading to a recommendation score higher than $X$ deserved. For instance, item popularity affects the exposure probability of an item in a previous recommendation strategy and interaction probability due to user conformity. Due to ignoring item popularity, CF methods will assign higher scores to items with higher exposure in previous recommendation strategies, leading to over-recommendation, \ie popularity bias. In the causal terminology, this type of bias effect is termed as \textit{confounding bias}. Beyond confounding bias, another source of bias in recommendation is the gap between the observed interactions and true user preference matching. Some item attributes directly affect the status of interactions. 

\noindent\textbf{Causal recommendation modeling.} The key to eliminating bias effects lies in modeling the causal effect of $X$ on $Y$ instead of the correlation between them. In causal language, it means viewing $X$ and $Y$ as treatment and outcome variables, respectively. The causal effect denotes to what extent $Y$ changes according to $X$, \ie the changes of $Y$ when forcibly changing the value of $X$ from a reference status to the observed value. To estimate such a causal effect, it is thus essential to incorporate conventional causal inference techniques into recommender models.
Consider video recommendations on platforms like YouTube and Netflix. Here, $X$ represents user preference while $Y$ symbolizes users' click behaviors. In an ideal setting, users' click behaviors should be directly influenced by their preferences. However, at times, external factors like an enticing video cover (represented by $Z$) might introduce bias.

\subsection{The Issues of Data Missing and Data Noise in Recommender Systems}
\subsubsection{Data missing in recommender systems}
The data utilized in recommender systems is typically limited, which cannot cover all possible user-item feedback.
For example, a user has only rated a small ratio of clicked movies; or the user purchasing the camera is not recorded as having bought a camera lens and a roll film, which is intuitively reasonable.
Therefore, the obtained data cannot fully represent the users' interest, leading to sub-optimal results for existing recommendation methods.
First, the interaction data observed is constrained by the already-deployed recommendation policy of the recommender system~\cite{saito2020asymmetric}. Users can only interact with specific items if these items are exposed to them, which strongly correlates with the recommender system's intrinsic strategy.
 In addition, users may refuse to give feedback~\cite{wang2019doubly}.
 For example, on movie rating websites such as IMDB or Douban\footnote{https://www.douban.com}, users may only rate a few of the movies they have watched. Under this condition, it becomes more challenging to model users' interests. 
Besides, features of users and items can also be missing in real-world recommender systems due to the high cost of feature collection. 

\subsubsection{The necessity of causal inference for data missing} Some earlier approaches~\cite{steck2013evaluation,schnabel2016recommendations,thomas2016data} without causal inference were developed to address the data-missing problem. Steck~\cite{steck2013evaluation} computes prediction errors for missing ratings. 
Schnabel \textit{et al.}~and Thomas \textit{et al.}~\cite{schnabel2016recommendations,thomas2016data} consider weights for each observed rating based on the probability of collecting that record. 
However, these methods are limited by low accuracy and poor generalization ability.
 Causal inference actually provides the causal descriptions of how data is generated, which can serve as prior knowledge to data-driven models. As a result, the negative impact of data-missing issues can be alleviated, improving accuracy and generalization ability.

\subsubsection{Data noise in recommender systems}
The recommender systems highly rely on the historical user-item interaction feedback to model users' preferences and predict the interaction probability between the user and the unseen item; thus, the reliability of collected data is the basis of the effectiveness of recommender systems. 
However, the data collected in the real world may be noisy, \textit{i.e., incorrect}.
It is hard to detect and eliminate noisy interactions in traditional recommendation methods.
Mahony \textit{et al.}~\cite{Mahony2006detect} classified data noise into two categories: \textit{natural noises} and \textit{malicious noises}. Natural noise relates to the noise generated during the data-collection procedure by recommender systems, and malicious noise denotes the noise being deliberately inserted into the system. 

As for the natural noise, Li \textit{et al.}~\cite{li2019collaborative} discussed various reasons that lead to the noisy data in recommender systems. 
The major reasons include the inaccurate impression of the users themself and the error in data collection. 
Jones \textit{et al.}~\cite{jones2011comparisons} points out that users can hardly accurately measure their preferences, thus leading to mismatch between their preferences and final ratings. 
Cosley \textit{et al.}~\cite{cosley2003seeing} found that noisy data arises when users map their opinions into discrete ratings. 
Zhang \textit{er al.}~\cite{zhang2021counterfactual} argued that in some streaming applications, the conversion events may be delayed to the time when data is collected.
Thus the feedback of users may have not yet occurred, resulting in a large number of incompletely labeled instances and introducing noise to data. 
Some existing work~\cite{wang2021denoising,hu2008collaborative,lu2018between,wen2019leveraging} also pointed out the difference between the implicit feedback and users' actual satisfaction because of noisy interactions.
For example, in E-Commerce, many clicks do not lead to purchases, and a large portion of purchases finally get negative comments.
Implicit interaction data widely used in recommender systems nowadays is easy to become noisy because of 
 the inaccurate first impression of users. Since users are exposed to a flood of information in today's online services, users are very likely to have accidentally triggered feedback such as click-by-mistake.

 As for the \textit{malicious noise}, it is produced by adversary attackers of recommender systems.
 For instance, on user-generated platforms such as TikTok\footnote{https://www.tiktok.com}, some authors will create plenty of new accounts to rate their work with high scores, trying to earn over-exposure opportunities.
 In e-commerce websites such as Amazon, some adversary sellers may generate fake order records or positive comments on their products.

 \subsubsection{The necessity of causal inference for data denoising}

Many previous works have experimentally demonstrated the severity of data noise and its negative effects on recommender systems. Cosley \textit{et al.}~\cite{cosley2003seeing} showed that only 60\% of users will keep their rating to the same movie when they are asked to re-rate for it. Further experiments show that statistically significant MAE differences arise when exploiting CF models on the original rating data and new rating data. Amatriain \textit{et al.}~\cite{amatriain2009like} showed that the recommendation performance will be significantly affected under noisy data compared to the noiseless data, with a difference of RMSE of about 40\%. Wang \textit{et al.}~\cite{wang2021denoising} found through experiments on two representative datasets the performance of recommender system trained by noisy data  experienced a performance drop of 9.56\%-21.81\% \wrt Recall@20 and drop of  3.92\%-8.81\% \wrt NDCG@20, compared with the recommender system trained over cleaned data. 
Although existing work has confirmed the widespread existence of data noise, which reveals that we need to consider its impact during training recommendation models, existing solutions are a few.
Data noise can arise from various sources, such as limitations in data collection (e.g., inaccurate values in users’ questionnaire data) or during data preprocessing (e.g., crudely transforming feedback into simplified values, like converting continuous user watching durations into discrete positive/negative labels). Such noises present significant challenges in accurately discerning user preferences. Leveraging causal inference allows us to more effectively detect the presence of noise in interaction data or bridge the disparity between noisy training data and the expected clean testing data with the help of counterfactual learning and reasoning.

\subsection{Beyond-accuracy Concerns in Recommender Systems}\label{sec::weakness-3}

Traditional recommender systems are designed towards the major goal of achieving higher accuracy,
\ie click-through rate or conversion ratio, serving for the platform benefit.
Nevertheless, 
as recommender systems have become
fundamental information services in more and more aspects of daily life, these concerns are not just technical problems but also social challenges.

\subsubsection{Explainability}
The requirement of explainability for recommender systems refers to the need that we should understand why some items are recommended while others are not.
It helps build a bridge between users and recommendation lists for better transparency and trustworthiness.
Specifically, it can be divided into two categories, explainable recommendation model and explainable recommendation results.
Some existing work~\cite{Wang2018ExplainableRV,Chen2020TryTI,Zhang2014ExplicitFM} mainly took some item aspects to give explanations, which is concluded as the aspect-aware explainable recommendation. For example, Wang \textit{et al.}~\cite{Wang2018ExplainableRV} learned users' preferences on given aspects by factorization method to get the aspect-aware explanations.

\noindent \textbf{The necessity of causal inference.}
Despite their effectiveness to some extent, existing methods of explainable recommendation are still limited~\cite{Tan2021CounterfactualER}.
Specifically, the explanation is built on correlation. As mentioned above, roughly extracting correlations from the observed data without the support of causal inference may lead to wrong conclusions.
Furthermore, the explanations of the recommendation model require building explicit causal relations between the components of the recommendation model and the prediction scores.
Additionally, the explanation for recommendation results should fully consider how different decision-factors, \textit{i.e.}, cause, lead to users' behaviors, \textit{i.e.}, effect.
Thus, achieving explainability is tightly connected to causal inference.

\subsubsection{Diversity and Filter Bubble}
Filter bubble describes the phenomenon where people tend to be isolated from diverse content and information by online personalization~\cite{pariser2011filter}.
As a consequence, users are placed in a fixed environment where they can only encounter similar topics or information. Passe \textit{et al.}~\cite{2017Homophily} attribute this effect to homogenization, which means people's behavior and interest show consistency and convergence.

The recommender system is one of the main causes of the filter bubble due to the principle of
generating
recommendation lists by learning the similarity between users or items~\cite{2011The}, which inevitably leads to homogeneous recommendations.
Gabriel Machado Lunardi  \textit{et al.}~\cite{2020A} empirically analyzed the filter-bubble formation based on popular CF methods and algorithms for diversified recommendation. 
In terms of human nature, researchers found that people tend to pursue a comfort zone and stay with the opinions they are interested in or agree with~\cite{2014Does}. 
In the long term, the filter bubble will narrow people's views and radicalize their ideas. 
Thus, it is an urgent problem to break filter bubbles and improve recommendation heterogeneity.

\noindent \textbf{The necessity of causal inference.}
The biased feedback loop is one of the most critical challenges in addressing the filter bubble, as learning from biased data will exacerbate the homogeneity in recommendation exposure and further bias the collected data. 
Moreover, the accuracy-diversity dilemma is another challenge, which refers to the phenomenon where pursuing accuracy will lead to low diversity. 
Causal inference provides the opportunity to address these challenges.
First, causal inference can alleviate the bias or missing data in collected data, supporting the exploration of unseen data.
Second, the causal inference-enhanced model can utilize the causal relationships under user behaviors, understanding why users consume certain items. This can help recommend items outside the existing categories and meet user demands.

\subsubsection{Fairness}
Recently, the fairness of recommendations has gained significant attention. As we know, recommender systems operate as multi-stakeholder platforms, thus encompassing various aspects of fairness concerns, including both user-side and item-side~\cite{2017Multisided}.

The user-side fairness issue arises from the diverse fairness concerns among users. For instance, while some users may be predominantly worried about potential biases based on their gender, others might be more concerned about age-related biases~\cite{Li2021personalizedCF}. To foster trust in the recommender system, it's essential to address these user-side fairness concerns in a personalized manner. While some approaches~\cite{Hardt2016EqualityOO} have attempted to rectify these fairness challenges using association-based methods—aimed at eliminating statistical metric discrepancies between groups—research has shown these methods to be inadequate and lacking in certain areas~\cite{Khademi2019FairnessIA,Kusner2017CounterfactualF}. Notably, these association-based techniques often overlook the intricate relationship between objective features and model outputs. Conversely, a few studies have explored fairness through a causal lens, offering insights into how output variables evolve with changes in input~\cite{Zhang2018EqualityOO,2018Fairness}.

Item-side fairness, on the other hand, evaluates the equity in treatment of each item during the recommendation process. Biases may emerge due to the oversight of particular items or their attributes. Several existing solutions~\cite{Gruson2019OfflineET,Schnabel2016RecommendationsAT} have ventured into unbiased learning or heuristic ranking adjustments to rectify these biases.

\noindent \textbf{The necessity of causal inference.}
Tackling fairness issues is akin to hypothesizing in a counterfactual realm: Had a user not been part of a specific group, or had an item lacked a certain feature, would the recommendation outcomes remain unchanged? If not, what would these altered recommendations look like? This difference between the counterfactual and factual worlds forms the cornerstone of fairness evaluation in recommender systems. Hence, methods grounded in causal inference, particularly those employing counterfactual reasoning, offer a fresh and more comprehensive approach to enhancing recommendation fairness compared to their non-causal counterparts.

\vspace{0.2cm}
In short, we have systematically discussed the limitations of existing recommender systems and why causal inference is essential to address these limitations. In the following, we will introduce how these challenges can be addressed (at least partially addressed) by presenting the recent advances in the causality-enhanced recommendation.

\section{Technical Details of Existing Works of Causal Inference-based Recommender Systems}\label{sec::existing-works}

\begin{figure*}[t!]
\centering
\includegraphics[width=0.8\linewidth]{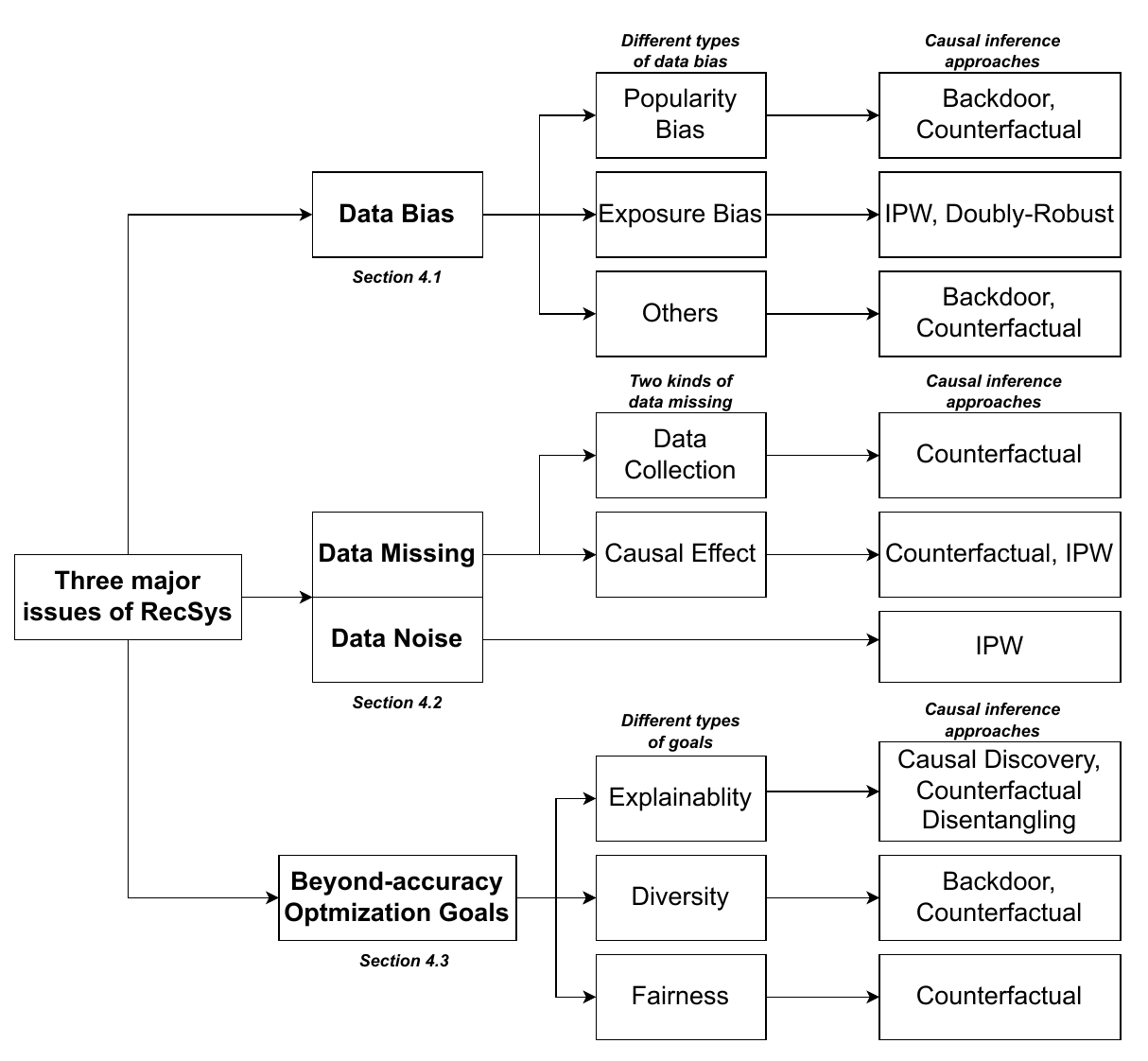} 
\caption{Illustration of existing work of causal inference for recommendation.} \label{fig:causal-recsys-methods}
\end{figure*}

The existing work of causal inference for recommendation is presented based on the three major issues of recommendation models with only correlation considered.
The overall illustration is presented in Fig.~\ref{fig:causal-recsys-methods}, and the details are introduced one by one as follows.

\subsection{Causal Inference-based Recommendation for Addressing Data Bias }

\begin{table*}[t]
    \centering
    \small
    \caption{Representative methods that utilize causal inference to address data bias .}
    \label{tab:bias}
    \begin{tabular}{c|c|c|c|c}
        \toprule
       \bf  Category & \bf Model & \bf Causal-inference Method & \bf Venue &\bf  Year\\ 
         \midrule
         \multirow{2}{*}{\bf Popularity Bias} & PD~\cite{DBLP:conf/sigir/ZhangF0WSL021} & Backdoor Adjustment & SIGIR & 2021\\
         & MACR~\cite{wei2021model} & Counterfactual Inference & KDD & 2021\\ \hline
         \bf Clickbait Bias & CR~\cite{DBLP:conf/sigir/WangF0ZC21} & Counterfactual Inference & SIGIR & 2021\\
         \bf Bias Amplification & DecRS~\cite{DBLP:conf/kdd/WangF0WC21} & Backdoor Adjustment & KDD & 2021\\
         \bf Conformity Bias & DICE~\cite{zheng2021disentangling} & Disentangled Causal Embeddings & WWW & 2021\\
 \bf 	\revise {Unknown Bias} & RD~\cite{ding2022addressing} & Doubly-Robust & KDD & 2022\\
 \bf 	\revise {General Feature Bias} &  DCR\cite{he2023addressing} & Backdoor Adjustment & TOIS & 2023\\
          \hline
         \multirow{10}{*}{\bf Exposure Bias} & IPS~\cite{Schnabel2016RecommendationsAT} & IPW & ICML & 2016\\
                  & MF-DR-JL~\cite{wang2019doubly} & DR & ICML & 2019\\
         & Multi-IPW/DR~\cite{zhang2020large} & IPW, DR & WWW & 2020\\
                  & Rel-MF~\cite{saito2020unbiased} & IPW & WSDM & 2020\\
         & DR~\cite{saito2020doubly} & DR & RecSys & 2020\\
         & MRDR~\cite{10.1145/3404835.3462917} & DR & SIGIR & 2021\\
         & LTD~\cite{10.1145/3437963.3441799} & RCT, DR & SIGIR & 2021\\
         & AutoDebias~\cite{10.1145/3404835.3462919} & RCT & SIGIR & 2021\\
         & USR~\cite{wang2022unbiased} & IPW & WWW & 2022 \\
         &DENC~\cite{li2023causal} & IPW & TKDD & 2023\\
                  \bottomrule
    \end{tabular}
\end{table*}

Existing methods on causal debiasing are mainly in three categories
: confounding effect, colliding effect, and counterfactual inference.

\subsubsection{Confounding Effect}
In most cases, biases are caused by confounders, which lead to confounding effect in correlations estimated from the observations. 
To eliminate the confounding effect, there are mainly two lines of research regarding the causal inference frameworks adopted.

\noindent \textbf{Structural Causal Model.}
Using SCM to eliminate confounding effect falls into two categories: backdoor and frontdoor adjustments. Backdoor adjustment is able to remove the correlations by blocking the effect of the observed confounders on the treatment variables. 
To address the data bias in recommender systems,
the existing work usually inspects the causal relationships in the data generation procedure, identifies the confounders, and then utilizes backdoor adjustment to estimate causal effect instead of correlation. 
Specifically, backdoor adjustment blocks the effect of confounders on the treatment variables by intervention \cite{pearl2009causality}, which forcibly adjusts the distribution of treatment variables and cuts off the backdoor path from treatment variables to outcome variables via confounders.

For example, Zhang \etal~\cite{DBLP:conf/sigir/ZhangF0WSL021} ascribed popularity bias to the confounding of item popularity, which affects both the item exposure and observed interactions. They then introduced backdoor adjustment to remove the confounding popularity bias during model training, and incorporated an inference strategy to mitigate popularity bias. 
Besides, Wang \etal~\cite{DBLP:conf/kdd/WangF0WC21} explored the bias amplification issue of recommender systems, \ie over-recommending some majority item categories in users' historical interactions. 
For instance, recommender systems tend to recommend more action movies to users if they have interacted with a large proportion of action movies before. 
To tackle this, Wang \etal~\cite{DBLP:conf/kdd/WangF0WC21} found that the imbalanced distribution of item categories is actually a confounder, affecting user representation and the interaction probability. Next, the authors proposed an approximation operator for backdoor adjustment, which can help alleviate the bias amplification.

However, the assumption of observed confounders might be infeasible in recommendation scenarios. To tackle the unobserved confounders (\eg the temperature when users interact with items), frontdoor adjustment is a default choice~\cite{pearl2009causality}. 
Xu \etal~\cite{xu2021deconfounded} has made some initial attempts to address both global and personalized confounders via frontdoor adjustment.
Zhu \etal \cite{zhu2022mitigating} gave a more detailed analysis of the conditions to apply the frontdoor adjustment in recommendation.
Liu~\textit{et al.}~\cite{liu2022practical} approached the selection bias challenge and proposed counterfactual learning-based method. Specifically, the authors focus on policy learning approaches for top-K recommendations in extensive item spaces, identifying key challenges like importance weight explosion and observation scarcity. A novel framework is introduced for efficient policy learning that addresses these complexities.
    Ding~\textit{et al.}~\cite{ding2022addressing} emphasizes the challenge of unmeasured confounders in recommender systems which can influence the accuracy of feedback predictions. The authors proposed Robust Deconfounder (RD) to consider the effects of these unmeasured confounders on propensities, using a bounded effect approach.

\noindent\textbf{Potential Outcome Framework.}
From the perspective of the potential outcome framework, the target is formulated as an unbiased learning objective for estimating a recommender model. Let $O^e$ denote the exposure operation where $o_{u,i} = 1$ means item $i$ is recommended to user $u$. The set $\mathcal{O}$ is defined as the exposure results under the given exposure strategy (with $\mathcal{O}^e$). According to the definition of IPW~\cite{little2019statistical}, we can learn a recommender to estimate the causal effect of $X$ on $Y$ by minimizing the following objective,
\begin{align}\label{eq:ipw}
    \frac{1}{|\mathcal{O}|} \sum_{(u,i)\in \mathcal{O}} \frac{ l(y_{u,i}, \hat{y}_{u,i})}{\hat{p}_{u,i}},
\end{align}
where $l(\cdot)$ denotes a recommendation loss and $\hat{p}_{u,i}$ denotes the propensity, \ie the probability of observing the user-item feedback $y_{u,i}$. 
As one of the initial attempts, Tobias \etal~\cite{Schnabel2016RecommendationsAT} adopted this objective to learn unbiased matrix factorization models where the propensity is estimated by a separately learned propensity model (logistic regression model). 
Beyond such shallow modeling of propensity~\cite{saito2020unbiased}, Zhang \etal~integrated the learning of propensity model and recommendation model into a multi-task learning framework~\cite{zhang2020large}, which demonstrates advantages over the separately learned one.
Wang~\textit{et al.}~\cite{wang2022unbiased} took the pioneering step of considering the exposure bias in the sequential recommendation, by proposing an IPW-based method named USR for alleviating the confounder in sequential behaviors.

Nevertheless, estimating the proper propensity score is non-trivial and typically suffers from high variance. To address these issues, a line of research~\cite{wang2019doubly,saito2020doubly,10.1145/3404835.3462917} pursues a doubly-robust model estimator by augmenting Equation~\ref{eq:ipw} with an error imputation model, which is formulated as:
\begin{align}\label{eq:ipw_dr}
    \frac{1}{|\mathcal{U}| \cdot |\mathcal{I}|} \sum_{(u,i)} \left(
        \hat{e}_{u,i} + 
        \frac{o_{u,i} (l(y_{u,i}, \hat{y}_{u,i}) - \hat{e}_{u,i})}{\hat{p}_{u,i}}
    \right),
\end{align}
where $\hat{e}_{u,i}$ is the output of the imputation model with user-item features as inputs. To learn the parameter of the imputation model, a joint learning framework~\cite{wang2019doubly} optimizes: \begin{align}
    \frac{1}{|\mathcal{O}|} \sum_{(u,i)\in \mathcal{O}} \frac{(l(y_{u,i}, \hat{y}_{u,i}) - \hat{e}_{u,i})^2}{\hat{p}_{u,i}}.
\end{align}
Undoubtedly, incorporating experimental data, \ie interactions from randomized controlled trial (RCT) such as random exposure, can enhance the doubly-robust estimator. In this light, a line of research~\cite{10.1145/3437963.3441799,10.1145/3404835.3462919} investigates data aggregation strategies, which largely focuses on tackling the sparsity issue of experimental data since RCT is costly.
Recently, Li~\textit{et al.}~\cite{li2023causal} considered that the exposure bias is largely depends on  the socially-connected users, and proposed IPS-based methods with the auxiliary social network data.

Different with the existing works for specific kinds of bias, He~\textit{et al.}~\cite{he2023addressing} studied how to address general feature biases.
This work identified a challenge in recommender systems where some features, like video length, can bias user interaction data and misrepresent actual preferences. Approaching from a causal perspective, the study introduced the Deconfounding Causal Recommendation (DCR) framework to address this bias. The DCR used backdoor adjustment to counteract the effects of confounding features and combine it with the mixture-of-experts (MoE) model architecture.

\subsubsection{Colliding Effect}

We can discover many collider structures (\cf Fig \ref{fig:DAGs}(c)) in the interaction generation process by inspecting the causal relationships. A representative case is that many different variables affect the observed interactions, such as user interests and conformity. Conditioning on the collected user interactions will lead to the correlation between user interests and conformity: an interaction caused by user conformity has a higher probability of being uninterested. To mitigate the conformity bias, an existing work \cite{zheng2021disentangling} disentangles the interest and conformity representations by training over cause-specific data, which improves the robustness and interpretability of user representations.

\subsubsection{Counterfactual Inference}
Another SCM-based technique used for debiasing is counterfactual inference.
In some SCMs of recommender systems, two causes (user and item features) lead to one effect (user behavior). If the user features or item features are significantly biased, this direct path (which we inaccurately referred to as a "shortcut" in our original version) can result in biased interaction learning, especially when other unbiased features take a more indirect route. 
The counterfactual inference is able to estimate the path-specific causal effect and eliminate the causal effect of partial user/item features. Specifically, it first imagines a counterfactual world without these features along specific paths and then compares the factual and counterfactual worlds to estimate the path-specific causal effect. For example, Wang \etal~\cite{DBLP:conf/sigir/WangF0ZC21} conducted counterfactual inference to remove the effect of exposure features (\eg attractive titles) for mitigating clickbait issues. In addition, Wei \etal~\cite{wei2021model} reduced the direct causal effect from the item node to the ranking score to alleviate popularity bias. Furthermore, Xu \etal \cite{xu2020adversarial} proposed an adversarial component to capture the counterfactual exposure mechanism and optimized the candidate model over the  worst-case scenario with a min-max game between two recommendation models.

\subsection{Causal Inference-based Recommendation for Addressing Data Missing and Noise}

Data collected from recommender systems are usually scarce due to limited user engagement compared with the whole item candidate pool.
In addition, the data can also be unreliable and incorrect since the system may fail to collect the true reward within the tight time window for data collection.
Meanwhile, the real causal effect of recommendation is largely unknown since the data of \textit{not recommending an item} is unavailable.
As a consequence, it is challenging for recommender systems to capture user preferences accurately since they are trained with missing and noisy data.
Tools of causal inference can be leveraged to tackle the two problems by generating either counterfactual data to augment insufficient training samples or counterfactual rewards to adjust noisy data.
Uplift modeling is utilized to measure the causal effect of recommendation.
Table \ref{tab:missing} provides a brief summary of recommender systems that utilize causal inference to address data missing and data noise problems.

\begin{table*}[t]
    \centering
    \small
    \caption{Representative methods that utilize causal inference to address data missing and data noise (RecSys refers to recommender system and CI refers to causal inference).}
    \label{tab:missing}
    \begin{tabular}{c|c|c|c|c|c}
        \toprule
         \bf Category &\bf Model &\bf   RecSys Task &\bf  CI Method & \bf Venue & \bf Year\\
         \midrule
         \multirow{13}{*}{\bf Data Missing} 
& ULO~\cite{sato2019uplift} & Collaborative Filtering & Uplift, IPW & RecSys & 2019 \\
         & DLCE~\cite{sato2020unbiased} & Collaborative Filtering & IPW & RecSys & 2020 \\
         & CauseRec~\cite{zhang2021causerec} & Sequential & Counterfactual & SIGIR & 2021 \\
         & CASR~\cite{wang2021counterfactual} & Sequential & Counterfactual & SIGIR & 2021 \\
         & CF\textsuperscript{2}~\cite{xiong2021counterfactual} & Feature-based & Counterfactual & CIKM & 2021 \\
         & CPR~\cite{yang2021top} & Collaborative Filtering & SCM, Counterfactual & CIKM & 2021 \\
         & CBI~\cite{sato2021online} & Collaborative Filtering & Interleaving, IPW & RecSys & 2021 \\
         & CausCF~\cite{xie2021causcf} & Collaborative Filtering & Uplift, RDD & CIKM & 2021 \\
         & DRIB~\cite{xiao2022towards} & Collaborative Filtering & Doubly-Robust, IPW & WSDM & 2022 \\
         & COR~\cite{wang2022causal} & CTR & Counterfactual& WWW & 2022\\
                  & CausPref~\cite{he2022causpref} & Collaborative Filtering & Causal Discovery & WWW & 2022 \\
                   & ASCKG-CG~\cite{mu2022alleviating} & KG-based & Counterfactual & SIGIR & 2022 \\
         &CIRS~\cite{gao2023cirs} & Sequential Recommendation & Counterfactual & TOIS & 2023 \\
         \hline
         \bf Data Noise & CBDF~\cite{zhang2021counterfactual} & Streaming & Importance Sampling & SIGIR & 2021 \\
      
         \bottomrule
    \end{tabular}
\end{table*}

\subsubsection{Causal Inference for Data Missing}
Interactions between users and items are the \textbf{factual} data, which expresses what really happens on the recommendation platforms and directly reflects user interest.
However, factual data is usually scarce; thus, it is insufficient for recommender systems to accurately capture the user interest hidden in the data.
The natural idea is to generate more samples that did not actually happen to augment the training data.
Such data augmentation aims to answer a question in \textbf{counterfactual} world: ``what would ... if ...'', which has been adopted in several research fields like computer vision~\cite{fu2020counterfactual}, and natural language processing \cite{zmigrod2019counterfactual}.
In terms of recommendation, counterfactual data augmentation aims to generate more interactions under situations that are different from the real cases when the factual data is collected.

\begin{figure*}[t!]
\centering
\includegraphics[width=0.9\linewidth]{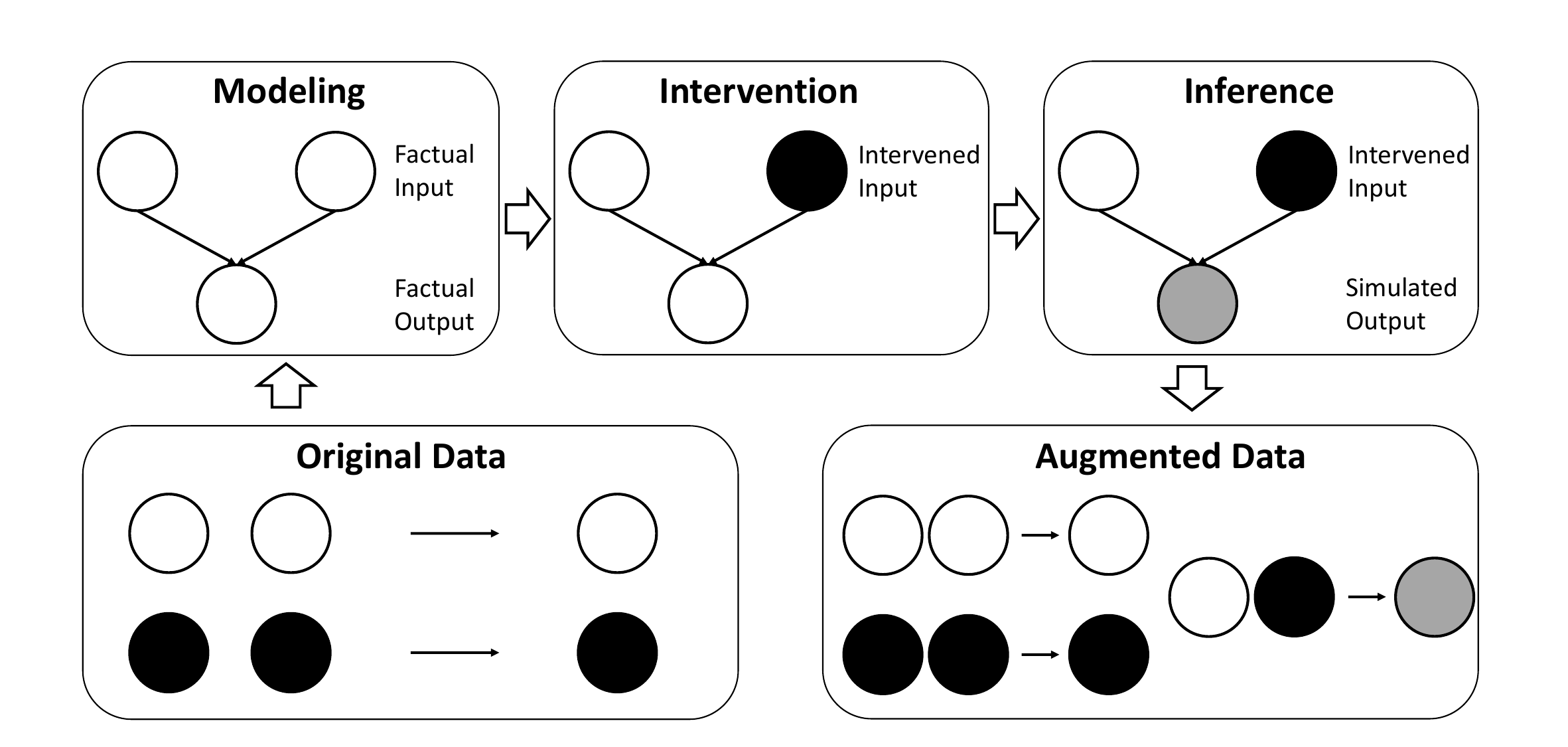} 
\caption{Illustration of counterfactual data augmentation for data missing.} \label{fig:data_missing}
\end{figure*}

Existing approaches answer counterfactual questions for the following recommendation scenarios,
\begin{itemize}[leftmargin=*]
    \item \textbf{Collaborative Filtering (Top-N Recommendation).} In this scenario, users are provided with a ranked list of items, and they will interact with several items in the list.
    Data augmentation generates the feedback of unseen recommendation lists; thus the counterfactual question is  ``what would the given user's feedback be if the system had provided a different recommendation list?''~\cite{yang2021top}. 
    \item \textbf{Sequential Recommendation.} In this scenario, recommendation is made according to the historical interaction sequences of users.
    In other words, interactions of the same user are regarded as a sequence ordered by the timestamp of each interaction.
    Augmented data are interaction sequences that do not exist in the real scenario.
    Therefore, the counterfactual question is ``what would users behave if their interaction sequences were different?''\cite{zhang2021causerec,wang2021counterfactual}.

    \item \textbf{Feature-based Recommendation.} In this scenario, not only interactions but also features such as user profiles and item attributes are available for recommendation.
    In other words, user preference modeling can rely on the user/item features.
    The counterfactual question that data augmentation aims to answer is ``what would the given user's feedback be if his/her feature-level preference had been different?''~\cite{xiong2021counterfactual}.

\end{itemize}
For all the above three scenarios, counterfactual data augmentation follows a similar paradigm of three steps, modeling, intervention, and inference.
Fig. \ref{fig:data_missing} provides a brief illustration of counterfactual data augmentation, and we will now introduce these three steps separately.

The \textbf{modeling} step captures the data generation process, which can be the recommendation model itself or another separate model.
Specifically, it is usually a parametric model that is trained to fit factual data.
In other words, given specific users and items that exist in factual data, the model serves as a \textbf{simulator} that is trained with the observed interactions and later generates unobserved interactions.
For example, Yang \textit{et al.}~\cite{yang2021top} first constructs a structural causal model to express the process of recommendation and then implements the SCM with an inner product between user and item embeddings.
Xiong \textit{et al.}~\cite{xiong2021counterfactual} utilizes a multi-layer neural network that takes feature vectors of users and items as input, then uses merging operators such as element-wise product or attention to fuse user and item feature-level properties.
Zhang \textit{et al.}~\cite{zhang2021causerec} and Wang \textit{et al.}~\cite{wang2021counterfactual} propose model-agnostic counterfactual data augmentation thus the model can be off-the-shelf sequential recommendation models.
The simulator is trained with existing factual data just as a normal recommendation task.
After a well-trained simulator is obtained, the input is intervened to be different from factual cases, and the simulator is used to produce the counterfactual outcome.
   Gao~\cite{gao2023cirs} studied counterfactual interactive recommender system (CIRS), which combines offline reinforcement learning with causal inference. The authors used a causal user model derived from historical data to understand the overexposure effect on user satisfaction, with which model, the RL policy can be better planned.

In the \textbf{intervention} step, the input is set as different values from the factual data.
Specifically, this step generates the counterfactual cases either by heuristic or another learning-based model.
Heuristic-based counterfactual intervention is usually achieved by randomization.
In~\cite{wang2021counterfactual}, a counterfactual interaction sequence can be generated by replacing an item at a random index with a random item.
In~\cite{zhang2021causerec}, dispensable and indispensable items are replaced with random items to construct counterfactual positive and negative sequences, respectively.
In contrast, the learning-based counterfactual intervention aims to construct more informative samples as data augmentation.
In other words, it generates counterfactual data with higher importance for model optimization.
For example, in~\cite{yang2021top}, a counterfactual recommendation list is generated by selecting items with larger loss value \textit{i.e.} the hard samples.
In~\cite{wang2021counterfactual} and~\cite{xiong2021counterfactual}, items and feature-level preferences that are at the decision boundary are selected and then modified with minimal change to construct more effective counterfactual interaction sequences and input features, respectively.

In the \textbf{inference} step, counterfactual outputs are generated with the above counterfactual inputs and simulator.
This step which uses the simulator to simulate the output of the intervened input, is usually straightforward.
In~\cite{yang2021top}, the counterfactual \textit{clicked} items of the intervened recommendation list are generated by inferring according to the constructed SCM.
In~\cite{zhang2021causerec} and~\cite{wang2021counterfactual}, the intervened interaction sequences are fed into the sequential backbone model, and the obtained outputs can directly serve as the counterfactual user embeddings~\cite{zhang2021causerec}, or they can be used to derive counterfactual next items~\cite{wang2021counterfactual}.

    Wang~\textit{et al.}~\cite{wang2022causal} further considered the problem of out-of-distribution recommendation, \textit{i.e.}, the data in another distribution is missing. The authors proposed to use a variational auto-encoder to help learn the user representations in the counterfactual distribution. Mu~\textit{et al.}~\cite{mu2022alleviating} proposed to use counterfactual generator to obtain user-item interaction data with the item's specific relation on the knowledge graph is changed. The counterfactual generator and recommender can be trained jointly to enhance each other.

A recent work~\cite{he2022causpref} approaches the issue of data-missing from another perspective, causal discovery.
	Specifically, this work delves into the vulnerability of current recommender systems to distribution shifts (the missing of IID data). The authors propose a novel causal preference-based recommendation framework named CausPref, integrating a recommendation-specific DAG learner.
	With emphasizing causal learning of invariant user preference and anti-preference negative sampling, CausPref shows superiority and interpretability in varied OOD settings.

\subsubsection{Causal Inference for Data Noise}

Interactions can be noisy or incorrect due to the tight time window of data collection.
For example, users' feedback can be delayed after the immediate interaction, such as purchasing an item a few days after adding it to the shopping cart.
In real-time recommendation, these samples are used for model training before the complete reward is observed.
Therefore, the reward at an early time is noisy, and whether the item will be purchased is unknown when it is added to the shopping cart.
Zhang~\textit{et al.}~\cite{zhang2021counterfactual} tackle the above problem of delayed feedback with the help of causal inference.
Specifically, the authors utilize importance sampling~\cite{bottou2013counterfactual,yasui2020feedback} to re-weight the original reward and obtain the modified reward in counterfactual world.

In addition, noisy user feedback can be alleviated by incorporating reliable feedback (\eg ratings). However, reliable feedback is usually sparse, leading to insufficient training samples. To solve the sparsity issue, Wang \etal~\cite{wang2021denoising} contributed a colliding inference strategy, which leverages the colliding effect~\cite{pearl2009causality} of reliable feedback on the predictions to facilitate the users with sparse reliable feedback.

\subsubsection{Causal Effect Estimation for Recommendation}
The recommender systems impact data collection, resulting in the absence of real interaction data, as mentioned above. Existing recommendation approaches are primarily evaluated and trained using interaction data, where typically, more interactions with recommended items indicate a more successful recommendation. However, they overlook the fact that some items may be interacted with by users even without a recommendation. Take e-commerce recommendation as an example: users may have clear intentions and directly purchase the items they desire. On the contrary, some items are more effective in terms of recommendation, meaning that users will purchase these items if recommended but won't purchase them if not recommended. Consequently, recommender systems boost the purchase probability of these effective items, referred to as \textit{uplift}. These items reflect a stronger causal effect of recommendation, emphasizing the importance of recommending more items with a larger uplift.

Some studies~\cite{sato2019uplift,sato2020unbiased,sato2021online,xie2021causcf} in recent years investigated the causal effect of recommender systems from the perspective of uplift.
Sato \textit{et al.}~\cite{sato2019uplift} applied the potential outcome framework to obtain the average treatment effect (ATE) of recommendation.
Specifically, all the interactions are divided into four categories according to the treatment (recommendation) and the effect (user feedback), and then a sampling approach named ULO is proposed to learn the uplift of each sample.
IPW was adopted to achieve unbiased offline learning~\cite{sato2020unbiased} and online evaluation~\cite{sato2021online} on the causal effect estimation of recommendation.
Xie \textit{et al.}~\cite{xie2021causcf} proposed to estimate the uplift with tensor factorization by regarding treatment as an extra embedding, and they use regression discontinuity design (RDD) analysis to simulate randomized experiments.
Xiao~\textit{el al.}~\cite{xiao2022towards} proposed a doubly-robust estimator, along with which a deep variational information bottleneck method is proposed to aid the adjustment of causal effect estimation.

Other studies view the causal effect of the recommendation algorithm as a problem related to off-policy evaluation. In reinforcement learning, the policy determines how the agent behaves (i.e., selecting the action) given the environmental context and the current states. In response, the environment provides the corresponding reward~\cite{levine2020offline}. However, due to high costs and limitations in data collection, it is challenging to collect all possible rewards for every action. Consequently, researchers have proposed off-policy evaluation, aiming to estimate these rewards~\cite{agarwal2017effective,sachdeva2020off}. In the context of recommendation, the items recommended can be viewed as the policy, and the off-policy evaluation is understood as estimating the effect of the deployed algorithm. This is akin to uplift modeling but focuses more on a general framework that estimates rewards using historical data. To achieve off-policy evaluation, there are three major categories of estimators: model-based estimators (reward regression), model-free estimators like propensity score-based methods, and hybrid estimators using doubly robust methods~\cite{saito2021counterfactual}.
Specifically, Swaminathan~\textit{et al.}~\cite{swaminathan2017off} tackled the problem of slate recommendation, where an ordered set of items is recommended. They built on techniques from combinatorial bandits to estimate a policy's performance using logged data. Li~\textit{et al.}~\cite{li2018offline} addressed a similar issue, aiming to estimate the number of clicks for a recommendation list. They introduced click models to construct estimators that learn with statistical efficiency, and the results showed the superior performance of these constructed estimators. Mcinerney~\textit{et al.}~\cite{mcinerney2020counterfactual} studied sequential recommendation and introduced a new counterfactual estimator that accounts for sequential interactions in the rewards, achieving lower variance. Specifically, they reweighted the rewards in the logging policy to approximate the expected sum of rewards under the target policy. Kiyohara~\textit{et al.}~\cite{kiyohara2022doubly} based their work on the assumption that users interact with items sequentially, starting from the top position in a ranking, leading them to propose a Cascade Doubly Robust estimator.

\subsection{Beyond-accuracy RecSys with Causal Inference}\label{sec::method::3}

\begin{table*}[t]
\small
    \centering
    \caption{Representative methods that utilize causal inference to achieve beyond-accuracy objectives (RecSys refers to recommender system and CI refers to causal inference).}
    \label{tab:accuracy-trap}
    \begin{tabular}{c|c|c|c|c|c}
        \toprule
         \bf Category & \bf Model &  \bf RecSys Task & \bf CI Method &\bf  Venue &\bf  Year\\
         \midrule
         \multirow{5}{*}{\bf Explanability}  & PGPR~\cite{xian2019reinforcement} &KG-enhanced&Causal Discovery& SIGIR & 2019 \\  
         & CountER~\cite{tan2021counterfactual}& CF &Counterfactual \& Causal Discovery&CIKM &2021    \\ 
        
         &MCT~\cite{tran2021recommending} &CTR&Couterfactual& KDD & 2021 \\  
         &CLSR~\cite{zheng2022disentangling} &Sequential&Disentangled Embedding& WWW & 2022 \\
         &IV4Rec~\cite{si2022model} & CTR & Decomposed Embeddings & WWW & 2022 \\
         \hline
         \multirow{3}{*}{\bf Diversity} & DecRS~\cite{DBLP:conf/kdd/WangF0WC21} & CF& Backdoor Adjustment & KDD & 2021\\

         & UCRS~\cite{wang2022user} &CTR&Counterfactual& SIGIR & 2022 \\
          &$\partial$CCF~\cite{xu2022dynamic} & CF & Backdoor Adjustment & CIKM & 2022 \\
         \hline
        \bf Fairness & CBDF~\cite{zhang2021counterfactual} &CTR&Counterfactual& SIGIR & 2021 \\
         \bottomrule
    \end{tabular}
\end{table*}

As mentioned in Section~\ref{sec::weakness-3}, non-causal recommender systems may find themselves focusing solely on improving accuracy, potentially overlooking other critical objectives such as explainability, fairness, diversity, and more.
In this section, we elaborate on how existing work addresses this challenge by introducing causal inference into recommender systems.

\subsubsection{Causal Inference for Explainable Recommendation}
Causal inference naturally can improve the explainability of recommendation, since it captures how different factors (cause) leads to recommendation (effect) rather than only the correlations. 
To present the existing works, we divide them into three categories as follows.

\begin{itemize}[leftmargin=*]
	\item \textit{Counterfactual learning.} Tan~\textit{et al.}~\cite{tan2021counterfactual} proposed CountER for explainable recommendation using counterfactual reasoning. CountER  explained the recommendation by highlighting the distinctions between factual and counterfactual scenarios. Specifically, CountER included an optimization task with the goal of identifying an item that minimizes the difference to the original item, thereby reversing the recommendation outcome in the counterfactual world.
CountER~\cite{tan2021counterfactual} also used causal discovery techniques to extract causal relations from historical interactions and the recommended items to enhance the explanation.

	\item \textit{Causal graph-guided representation learning.}
Zheng~\textit{et al.}~\cite{zheng2022disentangling} built a recommendation model based on the causal graph. The authors pre-define the causal relationships that how user behaviors (effect) are generated from users' two parts of preferences (causes), long-term preferences and short-term ones. Long-term preferences refer to those stable and intrinsic interests, while short-term preferences refer to dynamic and temporary interests.  The evolution manner is also defined for these two kinds of preferences.
Based on the pre-defined causal relations, the authors proposed to assign two disentangled embeddings for two parts of preferences, and the extracted self-supervised signals make the recommendation model explainable.
Si~\textit{et al.}~\cite{si2022model} proposed to improve the recommendation model's explainability by decomposing model parameters into two parts: causal part and non-causal part. Specifically, it built a model-agnostic framework by using users' search behaviors as an instrumental variable.

\item \textit{Causal discovery.}
Xian~\textit{et al.}~\cite{xian2019reinforcement} proposed to make use of a knowledge graph for explainable recommendation, and the paths in the knowledge graph can be used for generating explanations. For example, the reason for purchasing  AirPods may be that the user has purchased an iPhone before, and iPhone and AirPods are reachable in the knowledge graph via relation \textit{has\_brand} and node \textit{Apple Brand}.
Based on the knowledge graph and users' interaction history, the authors~\cite{xian2019reinforcement} proposed to extract causal relations by a reinforcement learning method. Specifically, the policy function of reinforcement learning is optimized to explicitly select items via paths in knowledge graph, ensuring high performance of both accuracy and explanation.
Tran~\textit{et al.}~\cite{tran2021recommending} approached the problem of explanable job-skill recommendation.
Specifically, it is essential to know which skill to learn to meet the requirements of the job. The authors first proposed causal-discovery methods based on different features with the employment-status label. Then the authors proposed a counterfactual reasoning method that finds the most important feature, of which the modification can lead to employment, which served as the explanations.
\end{itemize}

\subsubsection{Causal Inference for Improving Diversity and Alleviating Filter Bubble}
As mentioned earlier, focusing solely on accuracy gives rise to the issue of overly homogeneous content, resulting in the phenomenon known as the filter bubble.
By leveraging causal inference, which aids in gaining a deeper understanding and explicitly modeling the causal effects of user-decision factors, recommendations with improved diversity and the reduction of the filter bubble can be achieved.

\begin{itemize}[leftmargin=*]
	\item \textit{Counterfactual learning.} Wang~\textit{et al.}~\cite{wang2022user} proposed a causal inference framework to alleviate the filter bubble with the help of user control. Specifically, the framework allows users' active control commands with different granularity to seek out-of-bubble contents. Furthermore, the authors proposed a counterfactual learning method that generates new user embeddings in the counterfactual world to remove user representations of out-of-date features. By constructing counterfactual representations, the recommendation can keep both accurate and diverse.

\item \textit{Backdoor Adjustment.} Wang~\textit{et al.}~\cite{DBLP:conf/kdd/WangF0WC21}  approached the problem of homogeneous recommendation, by regarding imbalanced item distribution as a confounder between user embedding and the prediction score. Specifically, the authors used the backdoor adjustment to block the effect of the imbalanced item-category distribution in training data, partly alleviating filter bubble. The proposed method is model agnostic and thus it can be adapted to different recommendation models, including both collaborative filtering and click-through rate prediction.
Xu~\textit{et al.}~\cite{xu2022dynamic} employed a causal graph with loops to represent the dynamic recommendation process which leads to the filter bubble. A Dynamic Causal Collaborative Filtering ($\partial$CCF) model is proposed, which leverages back-door adjustment to estimate post-intervention user preferences and employs counterfactual reasoning to alleviate the echo chamber effect. Real-world dataset experiments validate the efficacy of the model in mitigating echo chambers, while maintaining strong recommendation performance.

\end{itemize}
\subsubsection{Causal Inference for Fairness in Recommendation} 
The concept of achieving fairness naturally aligns with the counterfactual world in causal inference. For instance, when evaluating the fairness of a recommender system for a specific user profile, one can pose a counterfactual question: \textit{Would the recommendation results change if the user profile were altered?}
Li~\textit{et al.}~\cite{Li2021personalizedCF} introduced the notion of counterfactual fairness in recommendation, where modifying the value of a given feature ensures that the distribution of recommendation probabilities remains unchanged.
The authors address this issue by introducing personalized fairness criteria for users. The core idea is to acquire user embeddings that are independent of specific features. To accomplish this, they propose a filtering module positioned after the embedding layer, which eliminates information relevant to sensitive features and generates filtered embeddings. Subsequently, the authors introduce a prediction module that utilizes these filtered embeddings to predict sensitive features, employing an adversarial learning approach in conjunction with the primary recommendation loss functions.

\section{Open Problems and Future Directions}\label{sec::open-problems}
We discuss important yet not-well-explored research directions in causal inference-based recommender systems.

\subsection{Causal Discovery for Recommendation}

We have systematically reviewed numerous works that integrate causal inference into recommender systems. However, existing approaches relying on predefined causal graphs or structural causal models exhibit two significant limitations.

 First, the assumed causal relationships may be inaccurate.  Although the recommendation tailored to the causal relations may improve the recommendation performance, hidden variables may exist that are the real causes.
Second, these manually crafted causal graphs are often simplistic, typically involving only a few variables, such as the user conformity, user interest, and user behavior in DICE~\cite{zheng2021disentangling}, the exposure feature, user/item/context features, and prediction score in CR~\cite{wang2021counterfactual}.
Nevertheless, users' decision-making processes may involve many factors in real-world scenarios. For example, whether a user visits a restaurant depends on the location, cuisine, brand, price, etc.
Therefore, it is essential to design causal discovery methods for learning causal relations from real-world data in recommender systems.
Traditional methods for causal discovery can be categorized into the following types. Constraint-based (CB) algorithms, such as the PC algorithm~\cite{spirtes2000causation} and the FCI algorithm~\cite{spirtes2001anytime}, initially identify conditional independence relationships between pairs of variables and then construct a directed acyclic graph based on these relationships. GES methods~\cite{chickering2002optimal,ramsey2017million} extend CB algorithms by incorporating a scoring function to assess the suitability of a directed acyclic graph (DAG). However, these established methods still grapple with challenges like high computational costs and limited robustness when dealing with large-scale data~\cite{guo2020survey}.
Recently, novel approaches based on deep learning~\cite{johansson2016learning,louizos2017causal,shalit2017estimating} and reinforcement learning~\cite{zhu2019causal} have emerged to infer causal relationships from extensive datasets. 
Therefore, it is a promising and crucial future direction for discovering causal relations and then leveraging the learned causal relations to enhance recommendation.

\subsection{Causality-aware Stable and Robust Recommendation}
Recommender systems are expected to be highly stable and robust, which can be explained in the following aspects.
First, the utilized data is dynamically collected, such as newly-registered users, new products, etc. As a result, the data distribution may be fast-changing~\cite{wang2022causal}.
Secondly, there exist multiple recommendation scenarios, including different tabs within the same mobile app, diverse domains, and various objectives. This necessitates that the recommendation model be capable of maintaining robustness and stability across these scenarios.
Last,  there exists a disparity between offline evaluations and online experiments. A recommendation model that performs well in offline experiments should ideally deliver strong results in  online environments.
In pursuit of greater stability and robustness in machine learning models, prior research~\cite{janzing2019causal,liu2020learning} has underscored the potential of causality-aware models. These models demonstrate a promising ability to adapt to different domains and excel in out-of-distribution (OOD) generalization~\cite{wang2022causal}.
Therefore, harnessing causality for the enhancement of robust and stable recommendations holds significant importance.

\subsection{Causality-aware Graph Neural Network-based Recommendation} 
In recent years, graph neural networks have been developing in recommendation at an unexpectedly fast speed.
GNN-based models have achieved strong performance in various recommendation tasks, such as the significant performance improvement of LightGCN~\cite{He2020LightGCNSA} against traditional neural network models~\cite{he2017neural} in collaborative filtering tasks.
The success of graph neural networks is mainly due to the strong ability to extract structured information, especially for the high-order similarity on the graph.
However, several critical challenges remain, awaiting solutions bolstered by causality.
First, there is a pressing need to demystify the workings of GNNs in making precise and successful recommendations. The explainability of powerful GNN-based recommendation models, encompassing both the model itself and the rationale behind recommendation results, remains an area ripe for further research. Currently, these models often operate as black boxes.
Second, while recent strides have been made in causality-aware recommendation models that incorporate GNN modules as integral components, the GNN module itself and the realm of causal inference remain somewhat separate. Explicitly intertwining the message-passing processes of GNNs with causal inference and reasoning for recommendation represents an open and uncharted research frontier.

\subsection{Causality-aware Simulator and Environment for Recommendation}
The recommender system is a kind of system that tries to estimate and recover how humans make decisions.
With a longer-term and more rational objective, such systems should not merely predict current or next-step user interactions but also take into account sequences of interactions, with the aim of maximizing user engagement or aligning with platform requirements.
Given the dynamic nature of user-system interactions, some prior works~\cite{ie2019recsim,shi2019virtual} have introduced simulators for recommender systems. These works specifically employ reinforcement learning techniques, including imitation learning~\cite{ho2016generative}, to simulate how users select items within specific environments and contexts.
However, these approaches are predominantly data-driven and often lack the underpinning of causality, potentially leading to inaccuracies in decision-making processes. Recently, causal reinforcement learning (CRL) methods have emerged to address the issue of missing data in reinforcement learning tasks.
Bareinboim~\textit{et al.}\cite{bareinboim2015bandits} introduced the concept of leveraging causal interventions to aid in estimating rewards while accounting for unobserved confounders. Additional works\cite{lattimore2016causal,yabe2018causal} have delved into causal bandit algorithms, offering theoretical bounds on performance improvements compared to non-causal bandits.
Causally-aware reinforcement learning approaches exhibit substantial promise in handling data limitations when modeling dynamic and sequential user-system interactions. Consequently, they are poised to play an indispensable role in modeling both the simulator and the environment of recommender systems.

\vspace{0.2cm}

\noindent To conclude, the future endeavors in the realm of causality-aware recommender systems should begin by addressing the constraints imposed by pre-defined causal graphs. Other promising avenues of research encompass enhancing robustness, which involves domain generalization, devising improved evaluation methods for long-term utility, bridging the gap between offline and online settings, exploring more effective integration with graph neural networks, and the development of causality-supported simulators for recommender systems.

\section{Conclusion}\label{sec::conclusion}

In recent years, causal inference has emerged as a critically significant and transformative topic within the realm of recommender systems research. Its significance cannot be overstated, as it has fundamentally altered our understanding of recommendation models.
This paper represents an initial stride towards presenting a comprehensive survey of existing literature in this domain. It meticulously and systematically delves into the rationale behind the applicability of causal inference and how it effectively mitigates the shortcomings inherent in non-causal recommendation models.
Our primary aim is to serve as a source of motivation for researchers already active in this field and, equally importantly, to inspire those who are contemplating the initiation of research endeavors in this exciting and burgeoning area.

\section*{Acknowledgment}
This work is supported in part by National Key Research and Development Program of China under 2020AAA0106000, and by National Natural Science Foundation of China under 62272262 and U23B2030. This work is also supported by grant from the Guoqiang Institute, Tsinghua University.

	\bibliographystyle{ACM-Reference-Format}
	\bibliography{bibliography}
\end{document}